\documentclass[aps,prb,superscriptaddress,twocolumn,tightenlines,showpacs,floatfix,amsmath,amssymb,pdftex]{revtex4}
\pdfoutput=1
\usepackage{graphicx}
\usepackage{subfigure}
\usepackage{color}
\usepackage{multirow}
\graphicspath{{./img/}{./}}

\usepackage{booktabs}
\usepackage{natbib}
\usepackage{hyperref}

\newcommand{\corr}[1]{\left<#1\right>}

\newcommand{\im}{i}

\newcommand{\doo}{\partial}

\newcommand{\de}{\mathrm{d}}

\newcommand{\real}{\mathbb{R}}

\newcommand{\integers}{\mathbb{Z}}

\newcommand{\set}[1]{\{#1\}}

\renewcommand{\Re}[1]{\textrm{Re }#1}

\newenvironment{matriisi2}{\left[\begin{matrix}}{\end{matrix}\right]}

\begin{document}

\title{
Classification of nematic order in 2+1D: Dislocation melting and $O(2)/Z_N$ lattice gauge theory}

\author{Ke Liu}
\affiliation{Instituut-Lorentz for Theoretical Physics,
Universiteit Leiden, PO Box 9506, NL-2300 RA Leiden, The Netherlands}

\author{Jaakko Nissinen}
\affiliation{Instituut-Lorentz for Theoretical Physics,
Universiteit Leiden, PO Box 9506, NL-2300 RA Leiden, The Netherlands}

\author{Zohar Nussinov}
\affiliation{Department of Physics, Washington University, St. Louis, MO 63160, USA}

\author{Robert-Jan Slager}
\affiliation{Instituut-Lorentz for Theoretical Physics,
Universiteit Leiden, PO Box 9506, NL-2300 RA Leiden, The Netherlands}

\author{Kai Wu}
\affiliation{Instituut-Lorentz for Theoretical Physics,
Universiteit Leiden, PO Box 9506, NL-2300 RA Leiden, The Netherlands}

\author{Jan Zaanen}
\affiliation{Instituut-Lorentz for Theoretical Physics,
Universiteit Leiden, PO Box 9506, NL-2300 RA Leiden, The Netherlands}

\date{\today}

\begin{abstract}
Nematic phases, breaking spontaneously rotational symmetry, provide for ubiquitously observed states of matter in both classical and quantum systems. These nematic states may  be further classified by their $N$--fold rotational invariance described by cyclic groups $C_N$ in 2+1D. Starting from the space groups of underlying $2d$ crystals, we present a general classification scheme incorporating $C_N$ nematic phases that arise from dislocation-mediated melting and discuss the conventional tensor order parameters.
By coupling the $O(2)$ matter fields to the $Z_N$ lattice gauge theory, an unified $O(2)/Z_N$ lattice gauge theory is constructed in order to describe all these nematic phases. This lattice gauge theory is shown to reproduce the $C_N$ nematic-isotropic liquid phase transitions and contains an additional deconfined phase.
Finally, using our $O(2)/Z_N$ gauge theory framework, we discuss phase transitions between different $C_N$ nematics.
\end{abstract}

%11.15.Ha	Lattice gauge theory
%64.70.dm	General theory of the solid-liquid transition
%Continuum models and theories of liquid crystal structure
%64.70.Tg	Quantum phase transitions (for quantum Hall effects aspects, see 73.43.Nq in electronic structure of surfaces, interfaces, thin films, and low dimensional structures)
\maketitle

\section{Introduction}
Among all exotic phases of strongly correlated electron systems, one of the most surprising is the electron nematic, \cite{Kivelson1998, Zaanen2004, Fradkin2010} a translationally invariant phase that breaks spontaneously rotational symmetry. \cite{Fradkin1999, Ando2002, Kee2003, Vladimir2006,  Fradkin2007, Fang2008, Xu2008, Vojta2009, Vojta2010} During the past two decades, experiments have proved the existence of such novel phases of quantum matter in strongly correlated electron systems such as High-$T_c$ superconductors\cite{Hinkov2008, Lawler2010, Mesaros2011} and parent compounds of iron-based superconductors,\cite{Chuang2010, Chu2012}  quantum Hall systems,\cite{Lilly19991, Pan1999, Lilly19992} and in the form of spinor/dipolar Bose condensation in optical lattices.\cite{Stamper-Kurn2013} The spin nematic \cite{Nussinov2002, Jan2003, Zhang2003, Tsunetsugu2006, Penc2011, Fernandes2014} has also been suggested as a candidate for the hidden order phase of the heavy fermion material URU$_2$Si$_2$.\cite{John2011, Fujimoto2011} Correspondingly, the classical liquid crystal theory developed by de Gennes\cite{deGennes} has successfully been extended to the quantum case to understand the physics of the quantum nematic with similar $D_{\infty h}$ uniaxial symmetry.\cite{Zaanen2004} One finds here an analogue in the form of the Pomeranchuk instability, conveying that the deformation of the fermi surface may be described by a tensor parameter similar to the one of the classical uniaxial nematic phase. It usually applies to the 2d nematic phase where the tensor order parameter can be further reduced to a scalar one, characterizing the anisotropy due to the rotational symmetry breaking, as has been already studied  extensively in the context of $2d$ electron liquid systems.\cite{Fradkin2010, Fernandes2014}  Most quantum nematic phases, however, occur in the (doped) strongly-correlated  Mott insulator, hosting a electron state reminiscent of the Wigner crystal.

Another route to nematic phases, developed by Zaanen and Kleinert, has been achieved by means of dislocation-mediated quantum melting of Wigner crystals. \cite{Kleinert2004, Zaanen2004, Vladimir2006} Here the condensation of dislocations effectively restores the translational symmetry of the crystal, while leaving the rotational symmetry broken. This is in essence an extension of the famous Kosterlitz-Thouless-Halperin-Nelson-Young (KTHNY)  theory,\cite{Kosterlitz1973, Halperin1978, Nelson1979, Young1979} describing the finite temperature hexatic phase resulting from topological dislocation melting of a triangular lattice,  to the quantum domain.

In the crystal phase, the continuous space symmetry is broken into a specific subgroup, which breaks both translational and rotational symmetry. This leads to the classification of crystalline lattices in both two and three dimensions, efficiently captured by the mathematical language of space groups.  As a result, when the translational symmetry is restored by dislocation condensation, there are different rotational symmetry subgroups descending from different space groups underlying the original crystals. Hence, there should be different nematic phases characterized by their invariance under different rotational subgroups, in addition to the uniaxial nematic with $D_{\infty h}$ symmetry. Despite examples like the classical $D_6$ hexatic in two spatial dimensions(2d) described by the KTHNY theory\cite{Kosterlitz1973, Halperin1978, Nelson1979, Young1979} and the quantum nematic with $C_2$ symmetry descending directly from the uniaxial nematic in 3d,\cite{Kivelson1998, Nussinov2002} the classification table of all nematic phases obtained in this fashion in both 2d and 3d  has not yet been provided.

The spatial dimension is critical when considering the broken rotational symmetry of the space groups and the resultant classification of nematic order, since the 2d rotational group $O(2)$ is abelian while the 3d rotations form a non-abelian structure. Hence, it is constructive to address the classification of nematic phases in the 2d abelian case and establish some basic principles that may be applicable to the non-abelian cases in 3d  for the further study.  To this end, we revisit the 2d case and provide the classification scheme of 2d nematic order, which allows us to establish an unified theory capturing all rotational symmetries and connect with all specific examples that were already extensively studied in Refs.\onlinecite{Lammert1993, Zaanen2004, Senthil2000, Vladimir2006}. Conventionally the phrase 'nematic' is reserved to phases with broken rotational symmetry by rod-like molecules which have  $C_2$ symmetry. The term 'hexatic' has then been invented to specify the nematic phase with $C_6$ symmetry. However, there is room for many different rotational symmetry broken phases, especially in the 3d case. This makes it tedious to specify every single phase individually. Moreover, all these phases break rotational symmetry in the same way and can hence can all be considered a ' nematic'. Therefore we will employ a systematic nomenclature to denote these phases. In particular, a 'nematic' phase with residual rotational symmetry $H$ ($H$ is the subgroup of $O(2)$ in 2d or $O(3)$ in 3d) is referred to as a $H$ nematic.
For example, one may consider generalizing the hexatic phase to a 3d $O_{h}$ nematic, which arises as a descendant from a $O_h$ cubic crystal by topological melting.

In this paper, we show that  dislocation condensation gives rise to five different classes of nematic phases invariant under different discrete subgroups $C_N$ of $O(2)$ with $N=1,2,3,4,6$. 
These nematics are therefore referred to as $C_N$ nematics, which correspond to the $p$-atic phases ($p=N$) identified by Park and Lubensky. \cite{Park1996}  Generalizing the $Z_2$ gauge theory in 3d \cite{Lammert1993, Lammert1995} describing the uniaxial nematic with $D_{\infty h}$ symmetry, we construct a general $O(2)/Z_N$ lattice gauge theory for all $C_N$ nematic phases in 2+1D by coupling a $O(2)$ matter field to a $Z_N$ lattice gauge field \cite{Kogut1979, Fradkin1979, Horn1979} with a nematic coupling $J$ and a defect coupling $K$. First, we comment on the symmetries and the construction of a general order parameter theory in two dimensions, making connection to earlier phenomenological proposals \cite{Park1996, deGennes}. By mobilizing the $Z_N$ gauge theory, we address the possible nematic phases and the associated phase transitions in terms of $J$ and $K$. This includes exotic $Z_N$ deconfined phases at large $K$, which may be related to exotic strongly coupled quantum phases. Analyzing the whole phase diagram, we first discuss  the conventional $C_N$ nematic-to-isotropic phase transition that arises in the small $K$ limit. In the $K\to\infty$ limit the partition function equates to that of the $XY$ model and we also discuss the large $K$ topological  $Z_N$ deconfined phase, which may be characterized by a string order parameter descending from the Fredenhagen-Marcu order parameter\cite{Fredenhagen1986,Fredenhagen1988}.  The gauge formulation allows us to discuss possible transitions between different $C_{N}$ nematics with considerable ease.

The remainder of the paper is organized as follows. In section II, we discuss the mechanism of dislocation proliferation in detail and present the full classification table of nematic phases obtained from dislocation-mediated melting in 2d. In section III, we then construct the corresponding order parameter $O(2)/Z_N$ lattice gauge theory for all $C_N$ nematic phases and discuss the corresponding phase diagram. In section IV, we focus on the strongly coupled limit of the $Z_N$ gauge field connecting the $O(2)/Z_N$ gauge theory to the conventional theory of nematic phases. Then in section V, we corroborate our previous results with Monte Carlo data. In section VI, we consider the emerging deconfined phase of the nematic gauge theory in detail and discuss the relevant string order parameter characterizing the topological order. In addition we show that the phase transition from the nematic to the deconfined phase belongs to the $XY^{\star}$ universality class. Finally, in section VII we comment on the possibility of phase transitions between different $C_N$ nematics within our gauge formalism.

\section{Crystalline dislocation-mediated melting}

In this section we consider the general classification scheme collecting the 2+1-dimensional quantum nematic phases that may arise as descendants from crystalline phases by topological melting. However, we do not directly address the existence of such a quantum melting transition from the parent crystal phase. For the general plausibility as well as the experimental realization of such a scenario, we refer the reader to the discussions in Refs. \onlinecite{Zoller2014, Nelson2014}.

In Section \ref{sec:OPK=0}, we then introduce the order parameters for the nematic phases and discuss the nature of the nematic-to-isotropic transition in terms of our symmetry classification.

\subsection{Melting picture}

The guiding principle in our classification is the central result stating that once the translational symmetries of the parent crystalline space group are modded out, one is left with the underlying point group of the crystal. In first instance it is immediately apparent that, due to the fact that the Burgers vector is fixed to the Bravais lattice, the condensation of dislocations leads to  a nematic phase breaking only rotational order as dictated by the Bravais structure. Nonetheless, the Burgers vector describing the dislocation is even more intricately tied to the crystal symmetry. In particular, the dislocation will have internal symmetry as imposed by the \textit{space group}, making up for a defect that only corresponds to translational symmetry.
Taking into account these 'sufficient conditions' in addition to the 'necessary' conditions set by the Bravais structure, we deduce the general classification table of 2d nematic order, showing that there are five $C_N$ nematic phases.

The point of departure is the observation that as a consequence of the structure of crystal symmetries, disclinations, conveying rotational order,
are massive and confined, once the translational symmetry is broken in the rotational plane \cite{Aron2013}.
This leads to the possibility of proliferating a system with dislocations, while the disclinations remain gapped. The process of
proliferation of dislocations then, in turn, effectively restores the translational symmetry and hence describes a
zero temperature crystal-nematic phase transition\cite{Kleinert2004}.  Due to the precise mathematical
description of the crystal symmetries in terms of space groups, this phase transition can
effectively be described with the respective symmetry.
Starting from the Euclidian group $E(2)$, the elements $\{A| \mathbf{t}\}$ of which transform a vector $\mathbf{r}$ by a rotation $A\in O(2)$ followed by a translation $\mathbf{t} \in \mathbb{R}^2$
\begin{equation}
\mathbf{r}\mapsto  A \mathbf{r}+\mathbf{t},
\end{equation}
a space group $G$ is a subgroup of the Euclidian group that has the property that the translations $ \mathbf{T}=\{ \mathbf{t}|\{I| \mathbf{t}\}\in G\}$ equate to a linear combination of primitive lattice vectors  $\mathbf{t}_{i}$. It is important to realize that $G/\mathbf{T}$  is isomorphic to the point group $P$. This essential property still holds for nonsymmorphic groups $N$, comprising point group elements $\{\{B|t\}|\{B|0\}\notin N\}$, as all translational symmetries are modded out.

These notions can then directly be employed to obtain the distinct nematic phases in 2d.
It is instructive to firstly consider the melting of a simple Bravais lattice, which is effectively obtained by applying $ \mathbf{T}$ to the origin. In such a structure it is particularly straightforward to visualize the effect of dislocations.
A dislocation is characterized by a Burgers vector, which represents the resultant vectorial lattice distortion.  The Burgers vector is fixed and can only equate to a linear combination of \textsl{primitive lattice} vectors and hence simply connects lattice sites of the original lattice, in the present case. As mentioned above, the condensation of many dislocations effectively destroys the long range translational order. However, in absence  of disclinations, the Burgers vector of each dislocation is a conserved quantity and the resulting phase thus still has rotational order, \textit{which is exactly captured by the point group of the original Bravais lattice.} We note that the point groups $D_{N}$ also contain elements describing the associated mirror symmetries. These additional constraints on the order parameter may be separated and are not considered in the remainder, as we are solely interested in the \textsl{rotational order}.  Henceforth, we can indicate the nematic phases by their characteristic  $C_{N}$ invariant. As a result, starting from the point groups underlying the Bravais lattices, it immediately follows that a  $C_2$, a $C_4$ and a $C_6$ nematic phase may be obtained from dislocation-mediated melting, see Table 1. In particular, the point group of the underlying Bravais lattice structure pertains to a 'necessary' condition for the rotational order of the nematic phases. It imposes the maximal symmetric rotational order resulting from the symmetry breaking of the Bravais structure, which may then be reduced by the full lattice symmetry.

\begin{figure}
\begin{center}
	\includegraphics[width=\columnwidth]{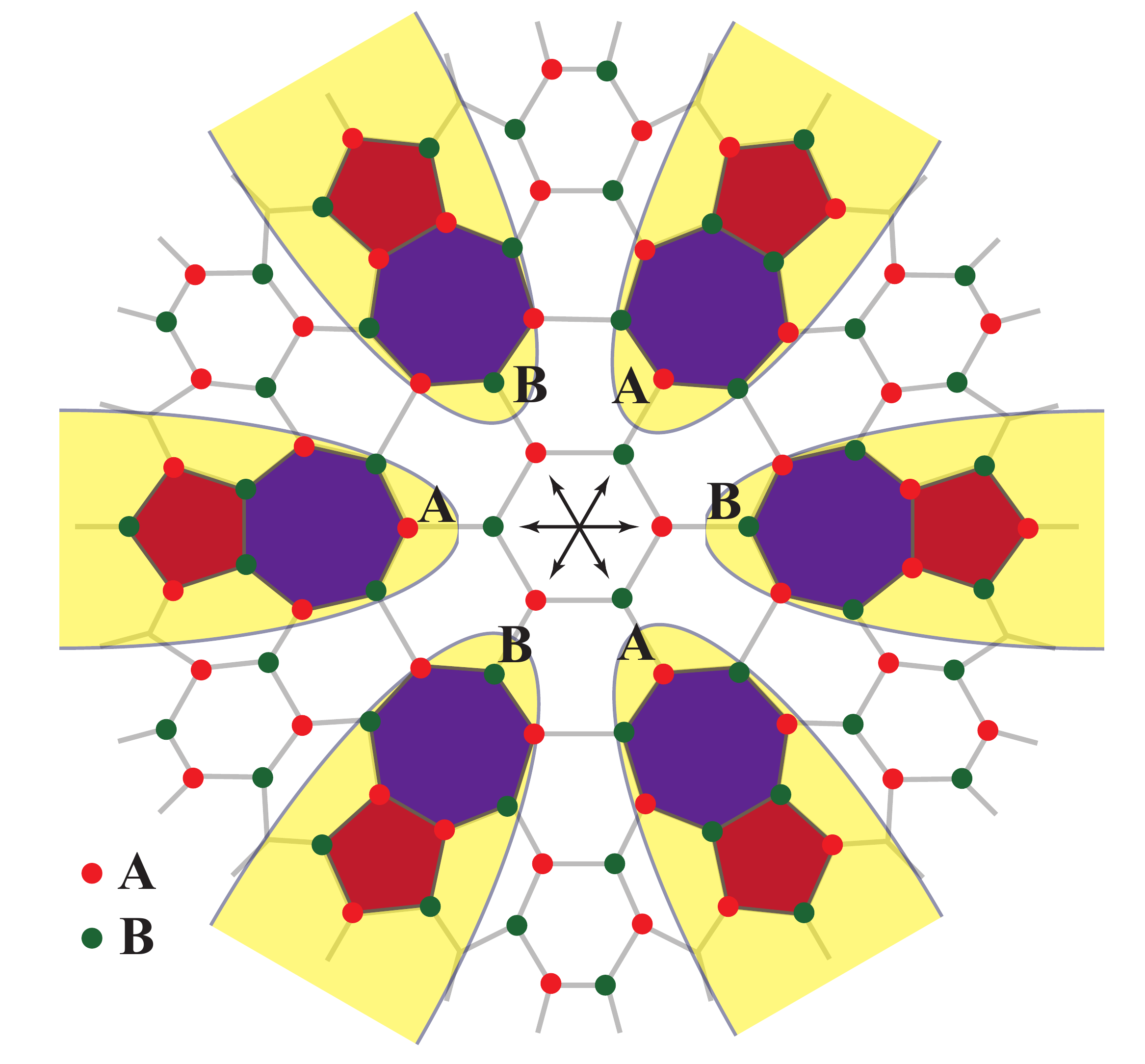}
	\caption{Figure displaying the dislocations in a A-B honeycomb lattice. The dislocation can be decomposed in a 'seven' and  'five' ring. On Bravais level, there are six types of elementary dislocations corresponding with the Burgers vectors (the directions of which are indicated with the black arrows in the center) oriented along the six different primitive lattice vectors. However,  as a $2\pi/6$ rotation maps A (B) onto the inequivalent  B(A), one obtains a $C_3$ nematic rather than a $C_{6}$ nematic.}
      \label{honeycomb}
\end{center}
\end{figure}

As a next step, the 'sufficient' conditions revealing the full classification of the nematic phases are then obtained by taking the space group into consideration.
We stipulate the fact that the dislocations are intimately tied to the translational structure of the crystal and therefore reflect the crystal symmetry encoded by the space group. For example, if the crystal symmetry is formed by multiple sublattices, the Burgers vector is still a primitive lattice vector, while the dislocation has internal structure as dictated by the translational symmetry.
Consequently, when the dislocations condense the symmetries of the unit cell are reflected via the underlying crystal symmetries at short range, whereas the collective nematic phase displays rotational order as revealed by the underlying point group. This may be illustrated by considering the representative example of  two inequivalent triangular lattices arranged into a honeycomb structure, see Fig. \ref{honeycomb}. As the Bravais lattice is triangular, one could naively argue that the Burgers vector can attain six distinct values, creating a $C_{6}$ nematic i.e. a hexatic. Crucially, however, the dislocation has an internal structure imposed by the space group, which breaks the sixfold rotational structure creating a $C_{3}$ nematic, connecting to the general statement that one should be left with the point group when the translational symmetry is effectively restored.
It is straight forward to apply this general procedure to any 2d space or so-called wall paper group.
As shown in Table 1, starting from the 17 space groups, this procedure leads to 5 different nematic states denoted as $C_N$, where $N=1,2,3,4,6$. We finally note that these classes of nematics are limited to the specific context of dislocation melting. In contrast,  there are quasicrystals with $C_5$ or $C_7$ symmetry. However, the dislocation melting mechanism for quasicrystals is still in the dark. Therefore, we exclude these cases and only consider nematic phases that can be obtained as descendants of real crystals.

\begin{table*}
\label{Table::class2D}
\begin{center}
\begin{tabular}{|c|c|c|c|}
\hline
Bravais Lattice Structure (PG) & $C_{N}$ Bravais Lattice & Space Group (PG) & $C_{N}$ Nematic Phase\\
\hline
\multirow{5}{*}{Hexagonal (D6)}&\multirow{5}{*}{$C_{6}$ }&p6mm $(D_{6})$ &\multirow{2}{*}{$C_6$ Nematic}\\
& &p6 $(C_{6})$ &\\ \cline{3-4}
& &p31m $(D_{3})$ &\multirow{3}{*}{$C_{3}$ Nematic}\\
& &p3m1 $(D_{3})$ &\\
& &p3 $(C_{3})$ &\\
\hline
\multirow{3}{*}{Square (D4)}&\multirow{3}{*}{$C_{4}$ }&p4mm $D_{4})$ &\multirow{3}{*}{$C_4$ Nematic}\\
& &p4gm $(D_{4})$ &\\
& &p4 $(C_{4})$ &\\
\hline
\multirow{5}{*}{Rectangular (D2)}&\multirow{5}{*}{$C_{2}$ }&p2mm $(D_{2})$ &\multirow{3}{*}{$C_2$ Nematic}\\
& & p2gm $(D_{2})$ &\\
& & p2gg $(D_{2})$ &\\ \cline{3-4}
& & pm $(D_{1})$ &\multirow{2}{*}{$C_{1}$ Nematic}\\
& & pg  $(D_{1})$ &\\
\hline
\multirow{2}{*}{Rhombic (D2)}&\multirow{2}{*}{$C_{2}$ }&c2mm $(D_{2})$ &$C_2$ Nematic\\ \cline{3-4}
& & cm $(D_{1})$ & $C_{1}$ Nematic\\
\hline
\multirow{2}{*}{Oblique (C2)}&\multirow{2}{*}{$C_{2}$ }&p2 $(C_{2})$&$C_{2}$ Nematic\\
\cline{3-4}
& & p1 $(C_{1})$&$C_{1}$ Nematic\\
\hline
\end{tabular}
\caption{The two dimensional nematic phases or $p$-atics which arise as descendants of crystals by topological melting. The first column shows the five Bravais lattice structures, with their corresponding point groups (PGs). Correspondingly, the second column displays the relevant $C_{N}$ group describing the rotational order associated with this Bravais lattice. The actual nematic phase is then obtained by considering the full space group and its associated point group, which may break the rotational symmetry to a smaller $C_{N}$ subgroup, as presented in the last two columns.}
\end{center}
\end{table*}

\subsection{Order parameter for nematic-to-isotropic phase transition}\label{sec:OPK=0} % and Landau free energy...?!
Having established the allowed symmetries of the $p$-atic phase ($p=N$) by dislocation melting, we now review the Ginzburg-Landau-Wilson order parameter theory, describing the $C_N$ nematic to isotropic liquid phase transition. The simplest orientational order parameter with a $C_N$ symmetry in two dimensions is the complex bond-order field\cite{Halperin1978, Nelson1979, Marchetti1990, Park1996}

\begin{align}
z_N(x) = e^{\im N\theta(x)},
\end{align}
where $\theta(x)$ is the angle of the orientational order parameter with respects to some fixed axis.

In the nematic ordered phase $\langle e^{\im N\theta(x)} \rangle \neq 0$, whereas in the isotropic liquid $\langle e^{\im N \theta} \rangle=0$. This immediately leads to the Ginzburg-Landau-Wilson action (in imaginary time formalism, see Appendix \ref{App:imaginarytime})

\begin{align}
S_{\textrm {eff}}=-J_{\rm{eff}}\int d^3x~(\partial_\mu z_{N}^* \partial_\mu z_{N}+c.c.), \quad z_Nz_N^* =1 \label{complex order} .
\end{align}
Here we emphasize that the action is supplemented with the constraint and that the physical order parameter is $z_N$ which has a well defined continuum limit at the transition. With this caveat, the universality class of the $C_N$ nematic transition is $XY$.

A representation in terms of real order parameters is obtained as follows. We can construct a corresponding $C_N$ invariant tensor order parameter from the effective action Eq.\eqref{complex order} by reassembling the imaginary and real part of $e^{\im\theta(x)}$ into a two-dimensional real vector $\vec{n}$ with and forming higher order tensors. To this end, following Park and Lubensky\cite{Park1996}, we introduce an $N$-rank complex tensor field for $z_{N}$:
 \begin{equation}
\mathbf{\Psi}_N=e^{i N \theta} \underbrace{\mathbf{\epsilon}_- \otimes ... \otimes \mathbf{\epsilon}_-}_{N \textrm{ times}} \label{base}
\end{equation}
where $\mathbf{\epsilon}_-=\frac{1}{\sqrt{2}}(\vec{e_1}-i\vec{e}_2)$ is a circular basis for the projection, in the sense that the rotor $\vec{n}$ can be expressed as $\vec{n}=\sqrt{2}\Re(e^{i\theta}\mathbf{\epsilon}_-)$ in this basis.

Rephrasing the effective action Eq.\eqref{complex order} in terms of the tensor bases $\mathbf{\epsilon}_- \otimes ... \otimes \mathbf{\epsilon}_-$, we obtain
\begin{align}
S_{\text{eff}}=-J_{\rm eff}\int d^3x(\partial_\mu Re\mathbf{\Psi}_N)^2+(\partial_\mu Im\mathbf{\Psi}_N)^2], \label{Eq:complexLandau}
\end{align}
where $Re\mathbf{\Psi}_N$ and $ Im\mathbf{\Psi}_N$ are $N$-order tensors contracted as
\begin{align}
(\partial_\mu Re\mathbf{\Psi}_N)^2=\partial_\mu (Re\mathbf{\Psi}_N)_{abc\cdots}\partial_\mu (Re\mathbf{\Psi}_N)_{abc\cdots}~.
\end{align}
This consideration is general since both $Re\mathbf{\Psi}_N$ and $Im\mathbf{\Psi}_N$ are symmetric for all pairs of indices, which makes different contractions of the tensors equivalent. Furthermore, an anti-clockwise $\pi/2$-rotation on $\epsilon_{-}$ just interchanges the real and imaginary parts of $\mathbf{\Psi}_N$, which are therefore redundant. This allows us to consider only \cite{Park1996}

\begin{equation}
Q_{N}=\sqrt{2}\Re\Psi_N  .
\end{equation}
where $Q_N$ is a traceless and symmetric $N$th rank tensor. Eq. \eqref{Eq:complexLandau} becomes
\begin{align}
S_{\rm eff}=-J_{\rm eff}\int d^3x(\partial_\mu Q_N)^2.
\end{align}
Note that in the case $N=2$, we retrieve the familiar order parameter for a $2d$ classical liquid crystal $Q_{ab}\sim (n_a n_b-\frac{1}{2}\delta_{ab})$. Similarly, for general $N$, a generalized $Q_{abc...}$ tensor can be obtained. For example for $N=3$, one gets
\begin{equation}
Q_3=Q_{abc}\sim n_a n_b n_c -\frac{1}{4}(n_a\delta_{bc}+ n_b\delta_{ca}+n_c\delta_{ab}).
\end{equation}
This order parameter can finally be employed (in a "soft spin" formulation) to obtain a Ginzburg-Landau-Wilson theory, which is nothing but a series expansion in powers of $Q_{N}$ in addition to lowest order gradients of $Q_N$. The allowed terms and coefficients are then as usual determined by a set of phenomenological parameters and global rotational invariance. Note that already for $N=2$ the term ${\rm tr} Q^3$ vanishes identically in two dimensions, and the transition is expected to be in the $XY$ universality class.

\section{Gauge theory description of quantum liquid crystals}

Let us now turn to an other route to address nematic ordering and phase transitions. In this scenario, instead of introducing a higher rank tensorial order parameter with the correct point group symmetries, one encodes the residual $C_N$ symmetry of the nematic by introducing gauged vectorial degrees of freedom, as in Refs. \onlinecite{Lammert1993, Lammert1995}. There the authors considered such a formulation especially fruitful since the symmetry of the order parameter as well as the role of topological defects are captured by the theory throughout the phase diagram. In fact, the gauge-defect term of their classical nematic leads to the possibility of a second order nematic-isotropic phase transition in three spatial dimensions. Apart from capturing the symmetries and the topological defects, our motivation for the gauge description of quantum nematics is also the possibility of strongly coupled quantum system with "emergent" nematic ordering and associated gauge fields. In this respect our approach is reminiscent of the so-called deconfined criticality scenario\cite{Senthil2004deconfined}. On the other hand, various realizations of quantum gauge-matter systems are relevant in quantum information theory \cite{Kitaev2006, Burrello2013}.

In general, the introduction of "fake" gauge symmetries is always allowed, since they merely represent redundancies in the full set of degrees of freedom in the theory. After fixing or eliminating the gauge degrees of freedom, the original physical variables are recovered. In particular, this applies to any physical observable, which are always required to be gauge invariant, as well as to any possible order parameter for a symmetry breaking phase transition of the orientational degrees of freedom, since it is impossible for a gauge non-invariant order parameter to develop a non-zero vacuum expectation value. In addition to correctly capturing the nematic degrees of freedom, the gauge formulation of the problem allows us to directly apply existing results available in the gauge theory literature. 

Since the symmetry group to be gauged is the discrete group $C_N \simeq Z_N$, the most straightforward approach is to define the theory on an auxiliary lattice. The resulting gauge theory, describing the $C_N$ nematic on a lattice, is given by $O(2)$ vector matter coupled to a $Z_N$ gauge field and will be referred to as $O(2)/Z_N$ theory in the remainder of this work. We note that it is the coupling to the gauge field that allows for the correct description of the $C_N$-nematic with only the residual $O(2)/Z_N$ orientational degrees of freedom. This is in essence a generalization of the $O(3)/Z_2$ theory used to describe the uniaxial nematic in three spatial dimensions \cite{Lammert1993, Lammert1995}.

To set the stage, let us first consider the $N=1$ case, which in our context could describe e.g. a ferroelectric nematic fluid \cite{Wei1992}. Obviously, the $C_1$ nematic is a special case since it is not invariant under any nontrivial discrete subgroup of $O(2)$. The effective theory for the orientational order is simply the $O(2)$ vector or $XY$ model in 2+1 dimensions. The $SO(2)\simeq U(1)$ vector/rotor $\vec{n}_i$ can be parametrized by a complex phase $n_i=e^{i\theta}$. As a result, the Euclidean action of the lattice theory in the imaginary time formalism takes the following form
\begin{align}\label{Eq:XY}
 S_{XY}=-\frac{J}{2}\sum_{\langle ij\rangle}(n_i^* n_j+c.c.) = -J \sum_{\langle ij \rangle} \cos(\theta_i -\theta_j),
\end{align}
where $J>0$ is the nematic (ferromagnetic) coupling on the regularization lattice. The ordered phase of the $XY$ model, with long range orientational order $\corr{n_i} = \corr{e^{\im \theta_i}} \neq 0$, then describes the $C_1$ nematic phase and the disordered rotationally invariant phase pertains to the isotropic liquid phase. The nematic-isotropic phase transition can be viewed as the profileration of topological defects, the $2\pi$-vortices of the $XY$ model.  These defects disorder the orientational order for $J < J_c$, the critical value of $J$, and finally lead to the liquid phase with the associated nematic rigidity $J\to 0$ at long distances.

\subsection{$Z_N$ lattice gauge theories for nematics}

\begin{figure}
  \centering
    \includegraphics[width=0.3\textwidth]{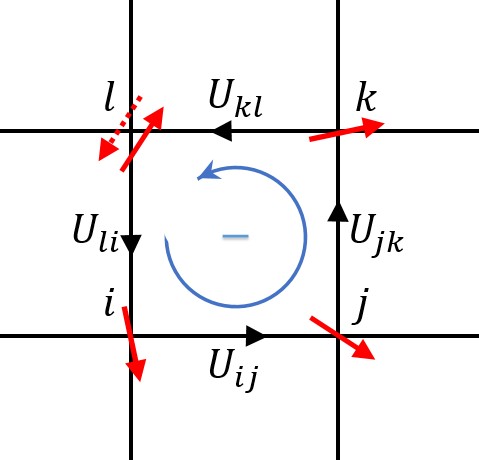}
  \caption{The conventions of the $XY-Z_N$ lattice gauge theory. The $SO(2)$ rotors $n_i$ are defined on lattice sites (red arrows). Gauge fields $U_{ij}$ are defined on the lattice links with orientations (black arrows). Note that the space-time lattice is an auxiliary lattice regulating the theory, while the point group symmetry of the nematic is reflected by introduction of the gauge fields. For example in the $Z_2$ case, the gauge symmetry makes reduces the original rotor at site $l$ in to a headless director (red and red dashed arrow). A disclination, represented by a frustrated plaquette (blue minus sign), is a configuration where the gauge links combine to a non-zero $Z_N$ vorticity of the rotor field when encircling the frustrated plaquette.}
\label{lattice}
\end{figure}

For the $C_{N\geqslant 2}$ nematic phases, however, a pure $XY$ model is not enough since it cannot reflect the symmetry of the orientational degrees of freedom in the ordered phase. Instead, we introduce a $Z_N$ gauge field that is minimally coupled to the rotors $n_i$. The gauge theory action can then be written as (in imaginary time, see Appendix \ref{App:imaginarytime} and e.g. Refs. \onlinecite{Kogut1979, Horn1979, Lammert1995})
\begin{align}\label{Eq:action}
S_{N}&=S_I+S_G
\end{align}
with
\begin{align}
S_{I}& =-\frac{J}{2}\sum_{\langle ij\rangle}(n_i^*U_{ij}n_j+c.c.)\label{matter action}
\\
S_{G}& =-\frac{K}{2}\sum_{\Box} \prod_{\corr{ij} \in \Box} U_{ij}+c.c. \label{gauge action}
\end{align}
The term $S_I$ is the lattice version of the minimal coupling of $n_i$ to a $Z_N$ gauge field $U_{ij}\in Z_N$, living on the lattice links, and $J > 0$ denotes the nematic (ferromagnetic) interaction. $S_G$ represents the simplest gauge invariant action for the gauge field $U_{ij}$, where $K$ is a coupling constant related to the gauge field strength, and the symbol '$\Box$' represents the elementary plaquettes of the cubic lattice composed of four nearest neighbor links.

Similar to the $XY$ model, the rotor field is represented as the complex phase $n_i=e^{i\theta_i}$ and the $U_{ij}\in Z_N$ can then simply be parameterized by a $U(1)$ phase: $U_{ij}=e^{-i\varphi_{ij}}$ with $\varphi_{ij}=2m_{ij}\pi/ N$, with $m_{ij} =1,2,\cdots,N-1$. Our orientation conventions are as shown in Fig \ref{lattice} and note that $U_{ji} = U^{-1}_{ij}$. As a result, the action Eq. \eqref{Eq:action} can finally be rewritten as
\begin{align}
S_I&=-J \sum_{\langle ij\rangle} \cos{(\theta_i-\theta_j + \varphi_{ij})}\label{Eq:XY-action}\\
S_G&=-K\sum_{ \Box}\cos({\varphi_{\Box}})\label{eq:Zn-action},
\end{align}
where we denote the lattice curl of $\varphi_{ij}$ as $\varphi_\Box=(\sum_{\langle ij\rangle \in\Box}\varphi_{ij})$. Written in this form, the action is clearly a generalization of the $XY$ model including additional gauge degrees of freedom. These are however only introduced to achieve the nematic point group symmetry, since the action $S_N$ is now invariant under arbitrary $Z_N$ gauge transformations
\begin{align}
  \theta_i&\rightarrow \theta_i+{2\pi\over N}\nonumber\\
  \varphi_{ij}&\rightarrow \varphi_{ij}+{2\pi\over N}\quad\text{for all adjacent links } \langle ij\rangle, \label{gauge transformation}
\end{align}
for each lattice site $i$.

Before going in to the details of the gauge theory and its phase structure, we will now first motivate the above form of the action as the description of the possible nematic order in 2+1 dimensions.

\subsection{Topological defects and gauge fields--- $Z_N$ disclinations in a nematic}
The form of the action $S_I$ in Eq. \eqref{Eq:XY-action} clearly reflects the point group symmetry $C_N\simeq Z_N$ in the orientational rotor field $\theta_i$. In contrast, the term $S_G$ with coupling $K$ has up to now only been justified by the fact that is allowed by symmetry. We will now show that it represent the elementary disclinations in the nematic and therefore plays a key role in the universal properties of the $C_N$ nematics.

The only nontrivial topological defects in the $C_N$ nematic phase are the disclinations. We expect that the phase transition and lack of nematic order is associated with the profileration of these defects. Due to the $Z_N$ symmetry, an elementary disclination is represented by a defect (Volterra-Frank) angle of $\theta_{\textrm{defect}} = \frac{2\pi}{N}$. Such an elementary $Z_N$ disclination can be constructed on the lattice as a gauge link configuration $\{U_{ij}\}$ satisfying
\begin{align}
\prod_{\corr{ij} \in \Box} U_{ij} = e^{2\pi\im/N} = e^{\im\theta_{\textrm{defect}}}\in Z_N,
\end{align}
around a particular plaquette $\Box$, since then the rotor field $n_i$ acquires a rotation of $\theta_{\rm defect}$ when encircled around $\Box$ in an anti clockwise fashion, see Fig. \ref{lattice}. Furthermore, clearly this defect angle is a gauge invariant property of the gauge field configuration $\set{U_{ij}}$.  In the imaginary time formalism, the gauge fields can be taken non-trivial only on spatial slices without loss of generality, leading to $Z_N$ "magnetic" fields. In fact, the gauge field allows us to construct all the defect angles representing disclinations of the point group $Z_N$ as configurations of the gauge fields $\set{U_{ij}}$. On the other hand, a full $2\pi$-vortex is captured by the configurations of the rotor angle $\theta_i$ as in the $XY$-model and does not require a non-trivial $Z_N$ gauge field configuration.

Inspection of the term $S_G$ now reveals that the extra gauge coupling $K$ represents a $Z_N$ defect (or a disclination or a vortex) suppression term. In fact, the role of $K$ is an effective disclination core energy, and is completely analogous to that appearing in \cite{Lammert1995}, or to the core energy appearing for the $2\pi$-vortices in the $XY$ model (usually parametrized in terms of the defect fugacity $y$).  Although one can of course assign different energies $K_i$ to the $N$ different disclinations in $Z_N$, we have for simplicity assigned the same coupling to all defect angles. The generalization will be briefly discussed in Section \ref{sec: deconfined phase}.

\subsection{Universal properties of the $O(2)/Z_N$ theory} \label{Sec:O(2)/Z_N to XY}
As shown below, the $O(2)/Z_N$ gauge theory Eq. \eqref{Eq:action} is characterized by a phase diagram that includes at least three phases: an isotropic liquid phase with disordered matter field and $Z_N$ gauge fields confined, a topological phase associated with deconfined $Z_N$ gauge fields, and finally an ordered nematic phase of matter field $n_i$, similarly as found in Ref. \onlinecite{Lammert1995}. Before turning to that discussion, we first want to describe how the gauge field and gauge symmetries are expected to affect the universal and critical behavior of the model as compared to the $C_1$ or $XY$ case. 

It is instructive first to consider the limit $K\to \infty$, where the $Z_N$ disclinations are completely suppressed. This sets
\begin{align}
\prod_{\corr{ij}\in\Box} U_{ij} = 1 \textrm{ for all $\Box$ in the lattice},
\end{align}
which allows us to write $U_{ij} = u_iu_j^*$ for $u_i = e^{2\pi\im m_i/N} \in Z_N$ without loss of generality (on a topologically trivial lattice). The resulting action is of the form
\begin{align}
S_{N}[K\to \infty] &= -J \sum_{\corr{ij}} \cos (\theta_i - \theta_j + \frac{2\pi(m_i - m_j)}{N})  \nonumber \\
&= -J\sum_{\corr{ij}} \cos (\theta'_i - \theta'_j),
\end{align}
which, by gauge symmetry, is just the partition function of the $XY$ model in the variables $\theta'_i = \theta_i +2\pi m_i/N$, and includes only $2\pi$-disclinations. For more detail, see the calculation in Appendix \ref{sec:Klarge}. In fact, this argument rigorously shows that by introducing the $Z_N$ gauge symmetry, the universal and critical properties are only affected for finite $K$, since the $K\to \infty$ partition function is that of the $XY$ model up to a irrelevant multiplicative constant coming from the gauge group volume (similarly as in e.g. the Mattis Ising spin glass \cite{Fradkin1978}).

We thus see that the full phase structure of the $C_N$ symmetric nematic is only revealed by also considering the role of the $Z_N$ disclinations appearing at finite $K$. A similar argument using gauge invariance and summing over the gauge transformations for any finite $K$ proves that only the gauge invariant content of the matter $\set{n_i}$ and gauge fields $\{U_{ij}\}$ is of relevance to the phase transition and universal properties (as e.g. only gauge invariant disorder or frustration is relevant in spin glasses \cite{Fradkin1978}). When combined with the well known triviality of all gauge non-invariant correlators (Elitzur's theorem), this line of arguments essentially completes the proof of the relevance of our gauge model Eq. \eqref{Eq:action} to describe the universal properties of nematic phases in 2+1 dimensions with $C_N$ point group symmetries.

\subsection{ $O(2)/Z_N$ phase diagram}
\begin{figure}[!tp]
\centering
\includegraphics[width=0.5\textwidth]{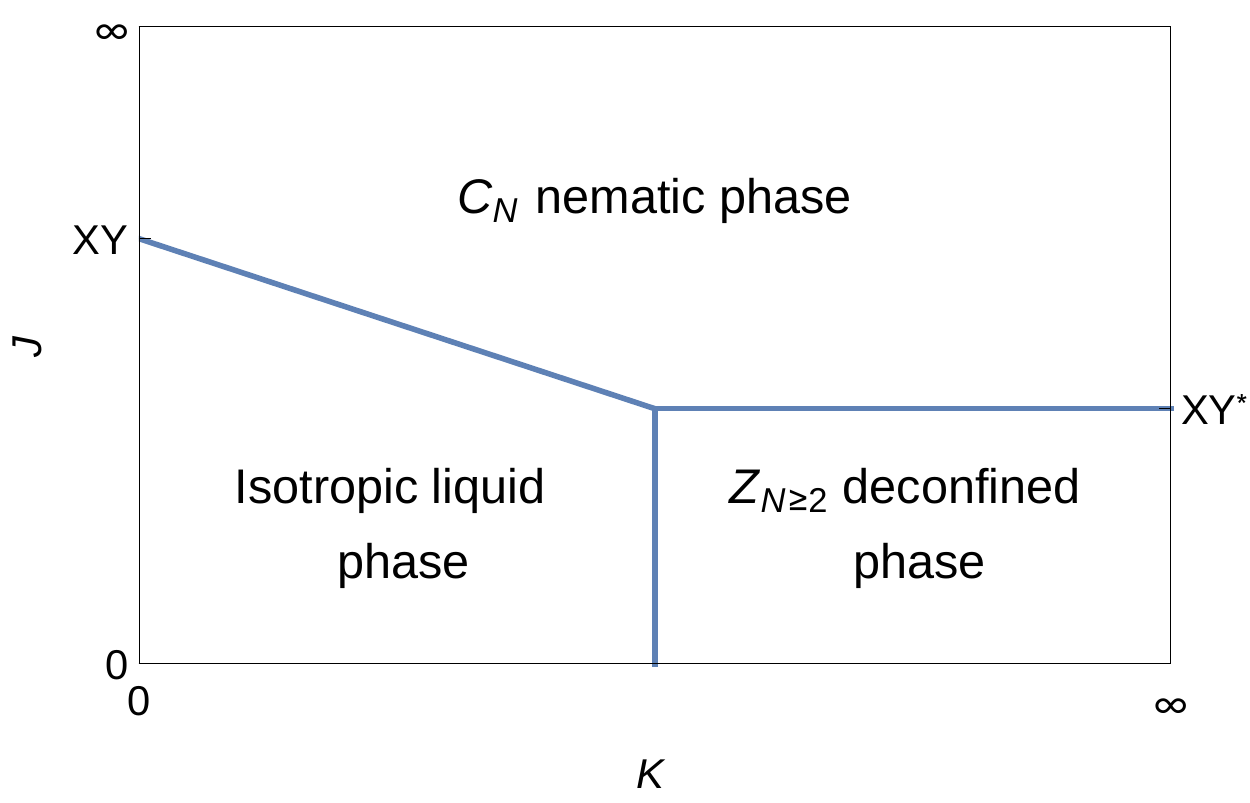}
\caption{ The schematic phase diagram of the $O(2)/Z_N$ gauge theory.
The $C_N$ nematic phase has long range orientational order and $Z_N$ disclinations. In the isotropic liquid phase at small $J$ and $K$, the $Z_N$ disclinations are condensed and the orientational order is destroyed. In the $Z_{N \geq 2}$ deconfined phase only $N$-tuples of $Z_{N \geq 2} $ vortices are condensed, leading to a phase with free $Z_N$ disclinations but no long range orientational order.}

\label{phase diagram}
\end{figure}

The $O(2)/Z_N$ effective theory contains XY-type rotor fields and $Z_N$ gauge fields, which both can go through phase transitions as a function of the couplings $J$ and $K$. The topology of the phase diagram is of course reminiscent of that in the $O(3)/Z_2$ lattice gauge theory \cite{Lammert1993, Lammert1995} and can be determined similarly by analyzing the phases appearing at suitable limiting values of $J$ and $K$. This results in the phase diagram shown schematically in Fig. \ref{phase diagram} which we now summarize.

{\it (i) $J\rightarrow0$ limit---} The matter becomes irrelevant and the theory describes a $Z_N$ lattice gauge theory with action $S_{G}$. The $Z_N$ gauge field undergoes a confinement-deconfinement phase transition as a function of $K$ \cite{Creutz1980, Borisenko2014}. In the confined phase for small $K$, the gauge field has large fluctuations leading to a condensate of the gauge fields that renders well-defined isolated fluxes absent. On the other hand, in the deconfined phase at large $K$, the flux excitations are gapped and isolated $Z_N$ fields exist in the spectrum. The characteristic behavior of the gauge field in these phases will also extend up to a region of finite $J$ \cite{Grover2010, Isakov2012, Sedgewick2002}. Moreover, the deconfined phase has topological order, characterized by a non-local order parameter. This will be discussed in Section \ref{sec: deconfined phase}.

{\it (ii) $J\rightarrow\infty$ limit---} For $J$ large (but finite), the $S_I$ term suppresses all  non-gauge fluctuations of rotor fields, since $\cos{(\theta_i-\theta_j-\varphi_{ij})}=0$, and thus the rotor and gauge fields are ordered, independent of $K$. Even for $K=0$, an excitation of the gauge flux is still gapped via the $S_I$ term and the spectrum contains Coulomb-confined neutral pairs of gauge excitations with finite energy. Hence, there is no phase transition as a function of $K$ for large enough $J$.  We identify this phase with ordered rotor fields and free $Z_N$ gauge flux excitations as the $C_N$ nematic phase.

{\it (iii) $K\rightarrow\infty$ limit---}
As shown in Sec \ref{sec:Klarge}, the partition function reduces to that of a regular $XY$ model. As a result of this equivalence, the system exhibits a three dimensional $XY$-type phase transition along the line $K = \infty$. However, the phase transition of the matter field is characterized by the $Z_{N}$ gauge invariant composite field $e^{i N\theta}$ rather than $e^{i\theta}$
and this effects the universality class of the transition. In the case $N=2$ this has been studied by various authors\cite{Grover2010, Isakov2012, Sedgewick2002} and was referred to as the $XY^{\star}$ universality class, which we will also adopt in the remainder.

{\it (iv) $K\rightarrow0$ limit---} Here the gauge fields do not have independent dynamics and the decrease of the nematic coupling $J$ drives a phase transition between the $C_N$ nematic phase and the isotropic liquid phase (the $Z_N$ confined phase with disordered rotors). On symmetry grounds this transition is expected to be in the $XY$ universality class and this will be discussed in more detail in Sec. \ref{Sec:K=0}.

\subsection{Dual description}\label{Sec:Dual}

There exist a well known dual formulation of the $XY$ model that emphasizes the role of the defects or vortices. This obtained by treating the matter field $\theta_i$ in the Villain approximation, as recollected in Appendix \ref{XYN Duality}. The dual formulation shows manifestly how the gauge symmetry encodes for the disclinations and their properties in the phase structure of our gauge model of nematics.

The corresponding dual action of Eq.\eqref{matter action} and Eq.\eqref{gauge action} can be written as
\begin{align}\label{eq: dual action}
\tilde{S}_{N} = -\frac{1}{8\pi^2J} \sum_{\tilde{\square}} A^{2}_{\tilde{\square}} - i \sum_{\langle \tilde{i}\tilde{j}\rangle (\Box) } A_{\tilde{i}\tilde{j}}(J^{XY}_{\Box} +\frac{\varphi_{\Box}}{2\pi}) \nonumber \\ -K\sum_{\square}\cos{(\varphi_{\square})},
\end{align}
where $A_{\tilde{i}\tilde{j}}$ is a non-compact $U(1)$ gauge field dual to the rotor field $\theta_i$. Here $\tilde{i}$ label the sites, $\langle \tilde{i}\tilde{j}\rangle$ the links, and $\tilde{\Box}$ the plaquettes in the dual lattice and are canonically associated, respectively, with the cubes, plaquettes and links of the original lattice. The dual gauge field strength is $A_{\tilde{\Box}}=\sum_{\langle \tilde{i}\tilde{j} \rangle\in \tilde{\Box}} A_{\tilde{i}\tilde{j}}$ and the original $Z_N$ gauge fields are $\varphi_{ij}={2\pi m_{ij}}/N$ with $m_{ij}$ integer mod $N$. 

Both the $XY$ vortices, represented by the integer current $J_{\Box}^{XY}$, and the $Z_N$ fluxes are charged under the dual gauge field $A_{\tilde{i}\tilde{j}}$, as in the normal $XY$ duality. However, from the second term in Eq.\eqref{eq: dual action} we see that the $Z_N$ vortices are fractionally charged and this leads to a statistical $Z_N$ phase that is attached to a flux of the $Z_N$ gauge field and to the flux of the dual gauge field $A_{\tilde{\Box}}$, which corresponds to the rotor current in the duality. Comparing this to the charge of the usual $2\pi$-vortices $J_{\Box}^{XY}$, we see that in the ordered phase of $O(2)/Z_N$ the $Z_N$-vortices indeed represent the $Z_N$-disclinations in the $C_N$ nematic.

Regarding the dual description of the phase structure we proceed as follows. Firstly, we see that the ordered phase at $J$ large is determined by the dual gauge field in the Coulomb phase of $U(1)$ gauge theory and the original $Z_N$ gauge symmetries. The phases with disordered rotor fields are characterized by (a Higgs) condensate of the $2\pi$-vortices $J^{XY}_{\Box}$ breaking the associated $U(1)$-gauge symmetry, just as in the usual $XY$ duality. In fact by referring to the coupling term in Eq. \eqref{eq: dual action}, the normal $XY$ transition can be considered as a condensation of $N$-tuples of $Z_N$ vortices. This effect, however, does not include the fractional $Z_N$ vortices and leaves an intact $Z_N$ gauge symmetry in the system for $K$ sufficiently large. At energies below the dual $U(1)$-photon mass gap, this disordered phase at large $K$ is non-trivial and described by the deconfined phase of pure $Z_N$ gauge theory with topological order. Similar $Z_q$ topological phases  appear in e.g. $U(1)$ gauge theory with $q$-charged matter \cite{Senthil2002,Hansson2004, Gregor2011} and also in a 3+1-dimensional compact $U(1)$-gauge theory with fractionalized flux lines \cite{Geraedts2014}, which is the generalization of the dual description Eq. \eqref{eq: dual action} of our $O(2)/Z_N$ model to higher dimensions. We will return to the detailed characteristics of the deconfined (topological) phase later in Section \ref{sec: deconfined phase}.

Finally, as $K$ decreases, also the $Z_N$ vortices can condense leaving no free gauge degrees of freedom describing a completely disordered and isotropic liquid phase. By the usual arguments of duality, the elementary excitations of this phase carry charges $2\pi / (2\pi/N) = N$ under the $Z_N$ gauge field, i.e. are necessarily gauge invariant.

\section{$K\rightarrow 0$ limit of the $O(2)/Z_{N}$ theory}\label{Sec:K=0}
In this section, we focus on the limit $K\rightarrow0$ of the $O(2)/Z_{N}$ gauge theory that features the nematic-to-isotropic phase transition. This bears most experimental relevance, as the deconfined phase for large $K$ is intimately related with the introduction of the gauge degrees of freedom and therefore contains auxiliary physics in addition to the nematic degrees of freedom.

We now consider the phase transition occuring in the $K\to 0$ limit. It is intuitively clear that when also $J$ is small, the rotor fields are disordered and the $Z_N$ gauge field is strongly fluctuating, describing an isotropic liquid phase. Increasing $J$ will then align the rotor field $n_i$ and drive a phase transition from the isotropic liquid to the $C_N$ nematic phase. In the $K\rightarrow0$ limit, the action Eq. \eqref{Eq:action} reduces to
\begin{align}\label{SK=0}
S_N[J,K=0]&=-J\sum_{\langle ij\rangle}(U_{ij}n_i^{*} n_{j}+c.c.) \nonumber\\
&= -J\sum_{\langle ij \rangle} \cos (\theta_i -\theta_j + \varphi_{ij}).
\end{align}
The $Z_N$ gauge fields $U_{ij}$ on different links are decoupled and can therefore be traced out to obtain an effective action for the matter fields. It is convenient to do so in the Villain approximation of the action in Eq. \eqref{SK=0} (see Appendix \ref{XYN Duality}),
\begin{align}
e^{-S_{\rm eff}} &= \sum_{\set{\varphi_{ij}}}e^{-S_N[J,K=0]}=\sum_{\{\phi_{i,\mu}\}} \prod_{\corr{i,\mu}} e^{J \cos (\triangle_{\mu}\theta_i+\phi_{i,\mu})} \\ 
&\to \sum_{\set{\varphi_{i,\mu}}} \sum_{\set{l_{i\mu}}\in \integers} \prod_{\corr{i,\mu}} 
N_V(J) e^{-\frac{J_V}{2}(\triangle_{\mu}\theta+\varphi_{i\mu}+2\pi l_{i\mu})^2}, \nonumber
\end{align}
where $N_V(J)$ is an unimportant (analytic) normalization factor. The sum over the gauge fields can now be reorganized as follows
\begin{align} \label{eq:Villain}
e^{-S_{\rm eff}} &\simeq \nonumber\\
\sum_{\{m_{i\mu}\}\in Z_N,\{l_{i\mu}\}\in \integers}& \prod_{\corr{i,\mu}} N_V(J) e^{-\frac{J_V}{2N^2}(N \triangle_{\mu}\theta+2\pi m_{i\mu}+2\pi N l_{i\mu})^2}\\
&= \sum_{\set{s_{i\mu}}\in \integers} \prod_{\corr{i,\mu}}N_V(J) e^{-\frac{J_V}{2N^2}(N \triangle_{\mu}\theta+2\pi s_{i\mu})^2} \nonumber\\
&= e^{-S_{V}[J_V/N^2]}\times N_V(J)^{N_l} \nonumber,
\end{align}
where $S_V[J']$ is the action of the $2\pi$-periodic Villain model with coupling $J'$. This model has a critical point at $J'_{c} \simeq 0.33$ in three dimensions \cite{Janke1986} and it follows that the model of Eq. \eqref{eq:Villain} is critical at coupling $J_{Vc} = J'_{c} N^2$. Using the relation between the original coupling $J$ in Eq. \eqref{SK=0} and the $J_V$ in the Villain model \cite{Janke1986},
\begin{align}
e^{-\frac{1}{2 J_V}} \simeq \frac{I_1(J)}{I_0(J)}, \label{Eq:VillainEstimate}
\end{align}
we obtain an estimate for the critical coupling $J_c(J_{Vc})$ of the $O(2)/Z_N$ model in the limit $K\to 0$. These values agree rather well with the critical coupling from our Monte Carlo simulations of the model in Eq.\eqref{SK=0}, as shown in Fig. \ref{Fig:J_c}. We further note that the gaussian model of Eq.\eqref{eq:Villain} itself of course also corresponds to a $2\pi$-periodic cosine model, however only in terms of the gauge invariant variable $N\theta_i$ as
\begin{align}
S_{\rm eff} \simeq -J_{\rm eff} \sum_{\corr{ij}} \cos N(\theta_i-\theta_j).
\end{align}
This is to be expected, since only gauge invariant terms appear after we have summed over the configurations of the gauge field. Close to the transition, this exactly reproduces the Ginzburg-Landau description of the nematic-to-isotropic transition in the $XY$ universality class.

\section{Monte Carlo results} \label{sec:PhaseK=0}
We have also simulated our gauge model using Monte Carlo in order to verify the topology of the phase diagram discussed above as well as the characteristics of the nematic-to-isotropic phase transition in the limit $K\to 0$.

\subsection{$K=0$ limit}
We simulated the gauge model for $K=0$ to verify that the transition is in the $XY$ universality class and check the Villain estimates Eq. \eqref{Eq:VillainEstimate} for the critical couplings $J_c$ as a function of $N$.

\begin{figure}[!tp]
\centering
\includegraphics[width=0.4\textwidth]{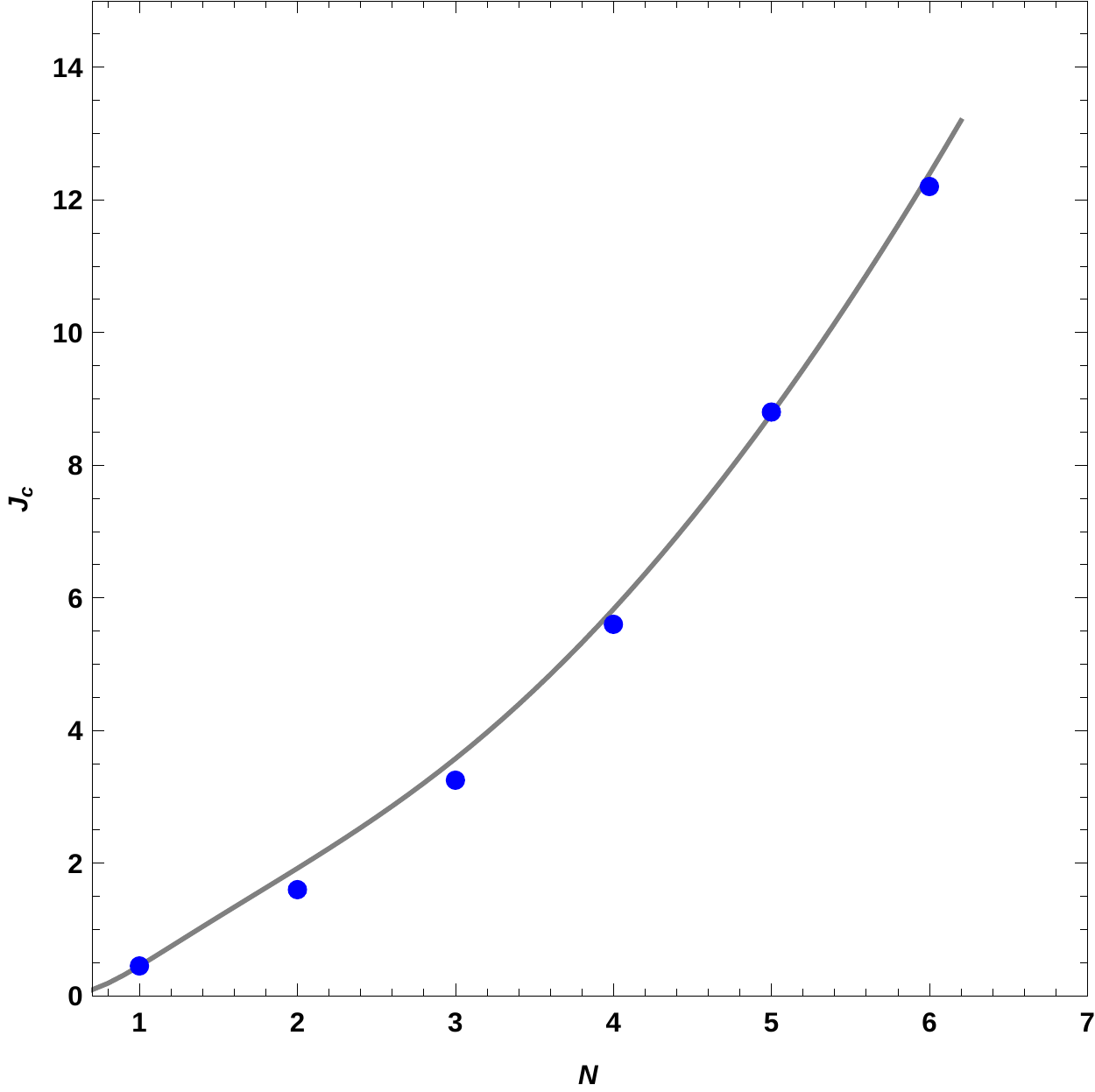}
\caption{The $K=0$ critical value $J_c$ of the nematic-to-isotropic transition as a function of $N$ from Monte Carlo data (blue dots). The line shows the critical coupling from the Villain estimate Eq. \eqref{Eq:VillainEstimate}.
}
 \label{Fig:J_c}
\end{figure}

To determine $J_{c}$ qualitatively at $K=0$, we employ standard Mote Carlo simulations, using the metropolis algorithm on a cubic lattice with $N_s=12^3$ sites with periodic boundary conditions. The obtained ensembles of equilibrium states were corroborated by comparing data obtained by heating ordered initial states at large $J$ and cooling disordered initial states at small $J$. The critical couplings $J_c$ for different $N$ can readily be estimated by computing the specific heat $C_V = \frac{1}{N_s}(\corr{S_N^2}-\corr{S_N}^2)$, local magnetization $m = \corr{|\Psi_{i}|}$ and susceptibility $\chi = N_s(\corr{m^2} - \corr{m}^2)$ for the gauge invariant quantity $\Psi_i = e^{\im N\theta_i}$, and associating developing singularities to a phase transition. The data for all $N \leq 6$ are consistent a transition in the $XY$ universality class. Our values for the critical couplings $J_c$ are shown in Fig. \ref{Fig:J_c} along with the values obtained from the Villain approximation.

\subsection{Phase diagrams}
\begin{figure}
\centering
\includegraphics[width=0.4\textwidth]{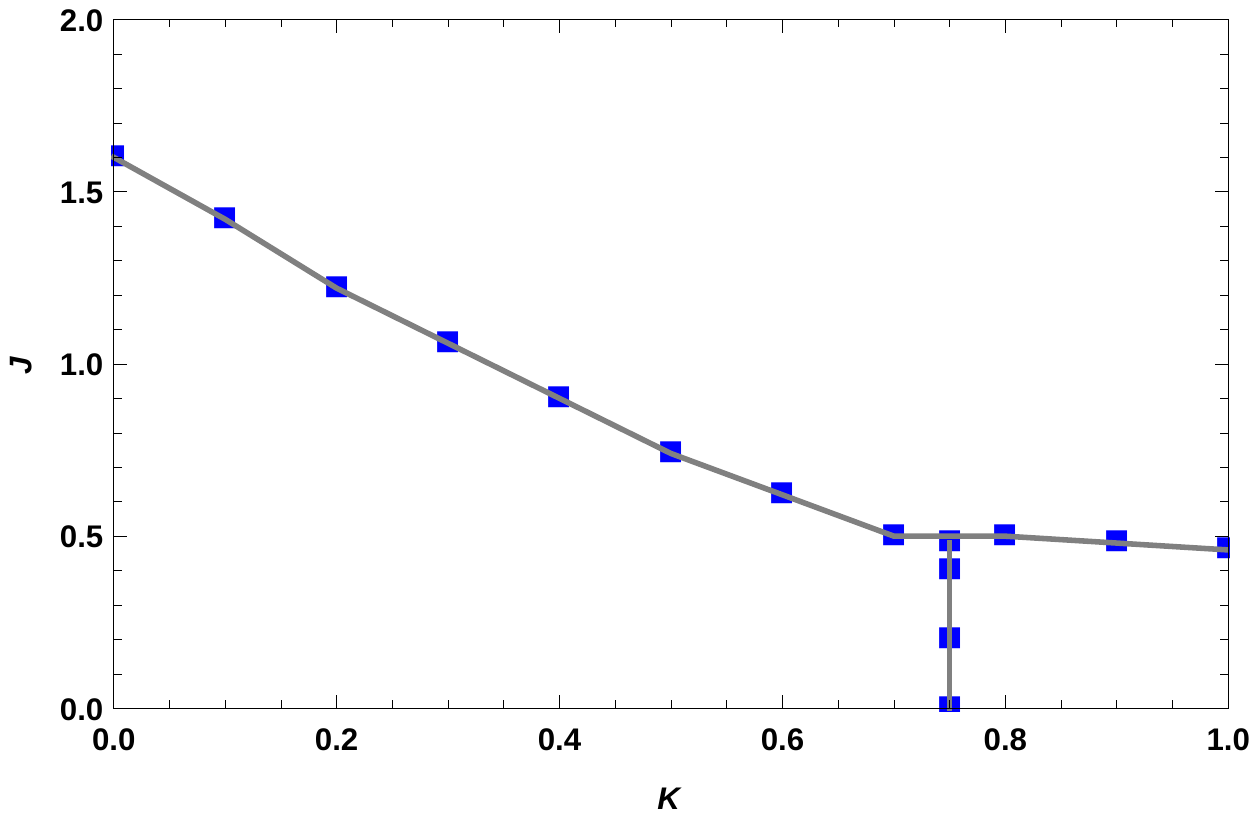}
\caption{The phase diagram of $O(2)/Z_2$ theory. The phases are identified as in Fig. \ref{phase diagram}.}
\label{Fig:Z2}
\end{figure}

\begin{figure}
\centering
\includegraphics[width=0.4\textwidth]{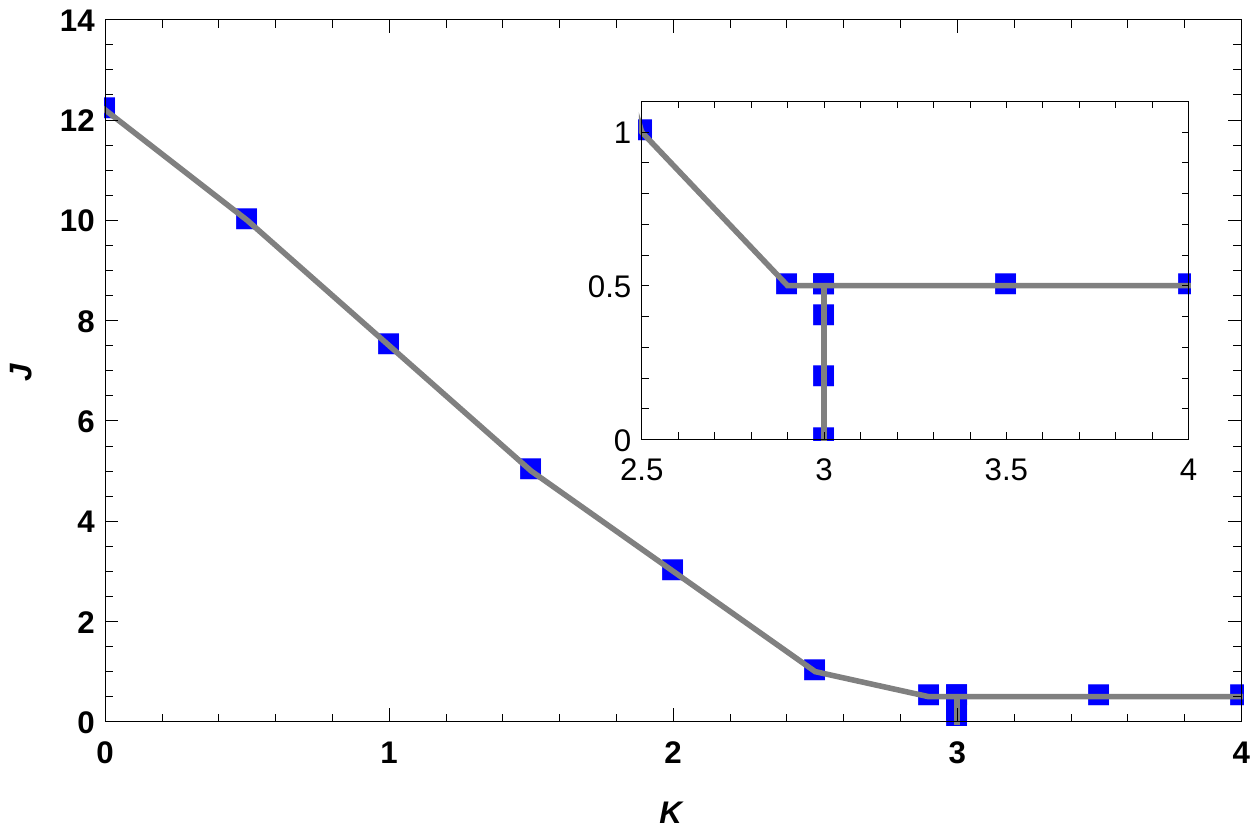}
\caption{The phase diagram of $O(2)/Z_6$ theory. The phases are identified as in Fig. \ref{phase diagram}, note the shrinking size of the deconfined phase (enlarged in the inset) as $N$ grows.}
\label{Fig:Z6}
\end{figure}

We have also simulated the gauge model \eqref{Eq:XY-action}, \eqref{eq:Zn-action} with Monte Carlo using the Metropolis algorithm on systems of size $N_s=12^3$ with periodic boundary conditions in order to verify the phase structure in the $K-J$ plane. To obtain the rough topology of the phase diagram, we monitored the peaks of the specific heat $C_V$ and the susceptibility $\chi$ and identified them with the critical values of the couplings $J_c,K_c$. Our results for the cases $N=2$ and $N=6$ are shown in Figs. \ref{Fig:Z2} and \ref{Fig:Z6}. As we have already noted, the critical value of $J_c(N)$ at $K=0$ grows roughly as $N^2$. On the other hand, the transition in the $K\to\infty$ limit is fixed at $J_c^{XY}\simeq 0.45$. The behavior of $K_c(N)$ as a function of $N$ for the pure gauge theory is also known \cite{Creutz1980}, with $K_c$ growing for larger $N$. This result to the fact that the size of the deconfined phase shrinks as a function of $N$, as is evident from Figs. \ref{Fig:Z2} and \ref{Fig:Z6}.

This comes as no surprise. First of all, when $N \rightarrow \infty$, the $XY$-$Z_N$ theory Eq. \eqref{Eq:action} tends to a $XY$-$U(1)$ theory, which is known to exhibit no phase transition for the  $K = 0$ line \cite{Fradkin1979}. Specifically, we can explicitly see that when $K=0$ in the strict $N \to \infty$ limit, the partition function becomes
\begin{eqnarray}
Z = \int \mathcal{D} [\theta_{i}] \int \mathcal{D} [\varphi_{ij}] e^{\sum_{\corr{ij}}J \cos (\theta_{i}-\theta_{j}-\varphi_{ij})} \nonumber
\\ = (2 \pi)^{N_{s} + N_{l}} I_{0}(J),
\end{eqnarray}
where $I_{0}$ is a modified Bessel function and $N_{s}$, $N_{l}$ are the number of sites and number of links respectively.
This function is analytic for all finite $J$. In fact, the partition function obtained is that of an $XY$ chain of length $(N_{s} + N_{l})$
which exhibits no phase transition for any finite $J$. Secondly, in the limit $N\to \infty$ the gauge group becomes a compact $U(1)$, and the whole line of phase transitions from the nematic to the isotropic liquid as well as the deconfined phase of the gauge theory disappear for any finite $K$\cite{Polyakov1975, Fradkin1979, Kleinert2002}, leaving a trivial phase diagram with no transitions.

\section{Beyond continuous rotational symmetry breaking}\label{sec: deconfined phase}
Although the existence of the $Z_N$ deconfined phase for large but finite $K$ is mainly an academic question when considering spatial rotational symmetry breaking, such a deconfined phase comes alive in the presence of internal rotational symmetry: A case in point being the spin degree of freedom.  In this section, we first enlighten the physics of the deconfined phase by discussing the connection between the $Z_N$ deconfined phase, stripe fractionalization, and the spin nematic phase (such as that of possible relevance to the high Tc superconducting cuprates).  We then discuss the topological nature and topological order parameter of the deconfined phase and the corresponding phase transition to the nematic.

\subsection{The deconfined phase}
The deconfined phase of our gauge model in Eqs. \eqref{Eq:XY-action}, \eqref{eq:Zn-action} is a phase exhibiting no long range order in the rotor fields due to the  proliferation of a {\em subset }of topological defects. We have already described this phase transition in terms of the dual formulation of our model in Section \ref{Sec:Dual}. To see how this happens in the original formulation, consider the defect structure implied by the different terms in Eqs. \eqref{Eq:XY-action}, \eqref{eq:Zn-action}. The $Z_N$ disclinations carry the core energy $\sim K$, whereas the $2\pi$-vortices have only an implicit core energy $\sim J$ coming from the cosine term of the $XY$-model. Therefore when $K$ is large, the $Z_N$ disclinations are gapped but at small enough $J \sim J_c^{XY}$, the $2\pi$-vortices will become gapless and proliferate. However, at the same time, the $2\pi$-vortices are $N$-fold tuples of $Z_N$ vortices and are favored energetically. Once the $XY$-vortices proliferate, the matter field will disorder and leave only the $Z_N$ gauge degrees of freedom. Since for large $K$ the gauge fields are deconfined, this phase is morally equivalent to the deconfined phase of the pure gauge theory \cite{WenQFT} and we will present a suitable string order parameter for this phase transition that involves both the matter and gauge fields.

For this phase to appear, we thus need conditions were it is possible to tune the nematic coupling $J$ and the core energy $K$ independently. In the context of a quantum nematic liquid crystal, this is basically equivalent to promoting the gauge fields to be independent degrees of freedom in addition to the orientational degrees of freedom. If the core energy of a single disclination (as described by the $K$ in the plaquette term) is very large, they can bind together to form $2\pi$-disclinations and liberate themselves from the $Z_N$ defect suppression. 
The subsequent proliferation of these $2\pi$ dislocations makes the system enter a non-trivial liquid phase without long range nematic order but free disclinations \cite{Lammert1995} that is described by the deconfined phase of the $Z_N$ lattice gauge theory.

To understand this phase, we can make an analogy to the spin nematic phase in the context of stripe fractionalization \cite{Jan2002,Jan2003}, where such physics is indeed encountered. Consider an antiferromagnet with long range stripe order in both charge and spin density.
Charge stripes act as domain walls or equivalently magnetic $\pi$--phase boundaries separating the Neel ordered regions.
In this case, the elementary topological defect of the stripe order is a spin dislocation on the bipartite lattice, carrying a half electric charge.
Such a defect causes spin frustration due to the bipartite antiferromagnetic order: the Neel vector changes direction when passing through a charge stripe. This resulting spin frustration can then effectively raise the core energy of the stripe dislocation.
However, this energy punishment can be evaded by binding two stripe dislocations into a double dislocation which is effectively a charge dislocation.
In the case of large energy cost per spin frustration, one thus identifies a scenario in which the stripe order is melted by the proliferating of only charge, i.e. pairs of the elementary dislocations.
The resulting phase is a stripe liquid exhibiting effective translational and rotational symmetry.
Nevertheless, this is an unusual liquid and in fact described by the deconfined phase of a $Z_2$ lattice gauge theory
where $Z_2$ vortices (visons) have the interpretation of stripe dislocations.
Such a spin nematic phase was first proposed by Zaanen et al. \cite{Jan2002,Jan2003} and further explored in Ref. \onlinecite{Podolsky2005,Kruger2002, Mross2012}. In particular in Ref.\onlinecite{Mross2012}, various stripe loop metal phases were studied.

\subsection{Topological order parameter and phase transition at large $K$}
Let us finally briefly comment on two characteristics of the deconfined phase: the topological string order parameter and the nature of the phase transition for the matter fields, relegating the details to the Appendix.

In addition to the  field $n_{i}=e^{iN\theta}$ that constitutes an order parameter for any phase transition driven by the coupling $J$ involving the $C_N$ ordered nematic, we need a topological order parameter for the gauge fields that is also adequate in the presence of charged matter fields and can identify the deconfined phase. It turns out that to this end we can define a string order parameter including {\em both} the rotor fields and  the gauge fields known as the the Fredenhagen-Marcu order parameter\cite{Fredenhagen1986,Fredenhagen1988}
\begin{align}
R(C_L)\equiv\frac{\mathcal{O}(C_{L/2})}{\sqrt{W(C_L)}}=\frac{\langle n^*_k\left(\prod_{ij\in C_{1/2}}U_{ij}\right)n_m\rangle}{\sqrt{\langle\prod_{ij\in C}U_{ij}\rangle}}.
\end{align}
In the above $C_{L/2}(k,m)$ is an arbitrary path of length $L/2$ connecting lattice sites $k$ and $m$ and $W(C_L)$ refers to a corresponding Wilson loop along a full loop $C_L$ of length $L$ with $C_{L/2}\subset C_L$.
It can be shown \cite{Fredenhagen1986, Fredenhagen1988, Gregor2011} that  this indeed distinguishes the $Z_N$ deconfined phase from the $C_N$ nematic phase and the isotropic liquid. Specifically,
\begin{align}
\lim_{L\rightarrow\infty}R(C_L)&=0 \quad \text{$Z_N$ deconfined phase}\nonumber\\
\lim_{L\rightarrow\infty}R(C_L)&\neq0 \quad \text{$C_N$ nematic or isotropic liquid}. \nonumber
\end{align}
The phase transition between the confined and deconfined phases in the pure gauge theory is captured by the Wilson loop, and the above can be considered as generalization of this in the presence of matter fields. 

Considering the phase transition in terms of the matter fields, the nematic-deconfined phase transition can be understood analytically in the $K\to\infty$ limit, as shown in detail in the Appendix. Namely, in this limit one can readily prove the equivalence of the matter coupled gauge theory of Eq. (\ref{Eq:action}) to that of the $XY$ model. This result actually not only holds for the gauge model with uniform coupling $J$, but also for the richer case of arbitrary couplings. One must, however, be aware of the caveat that the gauge invariant quantity $(n_i)^N=e^{iN\theta_i}$ is a composite field in the effective $XY$ model at $K\to \infty$. Specifically, the correlation function can be written as
\begin{align}
\langle (n_k^*)^N (n_m)^N\rangle_{K\to\infty}=\langle e^{iN(\theta_k-\theta_m)}\rangle_{XY} 
\end{align}
This affects some aspects of the universality class of the transition, for instance the anomalous dimension $\eta$ of the order parameter \cite{Grover2010, Isakov2012} and is usually referred to as the $XY^{\star}$ universality class. The qualitative features of this universality class are expected to carry over to finite $K$ up to the tricritical point, which is consistent with our Monte Carlo simulations, with the important addition of the phase transition of the gauge fields. We conclude that the phase transition is fully described by the topological string order parameter and the $XY^{\star}$ transition of the matter fields.

\section{Nematic phase transitions}\label{sec:par}

Nematic phases are usually analyzed within the Ginzburg-Landau-Wilson framework in terms of phenomenological continuum theories for the nematic degrees of freedom at the phase transition, as we discussed in Section \ref{sec:OPK=0} for the nematic-to-isotropic transition. Apart from this phase transition, a more interesting example of a nematic phase transition based on symmetry breaking is encountered by starting from a nematic with high symmetry and driving a phase transition to a nematic phase with lower symmetry.

In our context this means that the $C_N$ nematic phase can in principle also undergo a phase transition that breaks the $C_{N}$ symmetry to a lower subgroup. Here we point out how one can incorporate simple arguments within our gauge formalism to describe such phase transitions between different nematics that would be more involved in terms of phenomenologically constructed Landau-type theories.

\subsection{Phase transitions between different $C_N$ nematic phases}\label{Sec:nematic transitions}

Here we describe phase transitions between different $C_N$ nematics by additional matter fields with Higgs terms. Since the $C_N$ symmetry of the nematic is described by the $Z_N$ gauge symmetry, the addition of suitable Higgs terms is capable of "breaking" that gauge symmetry to a specific subgroup. We note that until now we have described the phase transitions by the condensation of the gauge defects, whereas the Higgs terms arise from non-trivial background fields, as in e.g. the stripe phases. For example, the hexatic phase with a $C_6$ symmetry can in principle break to a $C_3$ nematic, if there is a possibility to introduce the $A$ and $B$ sublattice inequivalence as shown in Fig.\ref{honeycomb}. 

In the $O(2)/Z_N$ gauge theory,  this phase transition can easily be accounted for in the $Z_N$ gauge sector.  Namely, it can be driven by an extra Higgs term in addition to the $Z_N$  gauge theory Eq.\eqref{gauge action}:
\begin{align}
      S_{Higgs}= -M\sum_{\corr{ij}} \sigma_i^{*}U_{ij}^{N/2}\sigma_j +h.c.  \label{Higgs}
\end{align}
where $\sigma_i$ is an Ising (or $Z_2$) field with charge $N/2$. When the Ising field is ordered $\langle \sigma_i\rangle \neq 0$, we can pick the unitary gauge where $\sigma_i \equiv 1~ \forall i$ and therefore the Higgs field completely drops out from the dynamics. However, despite the Higgs term, the theory still has a gauge symmetry given by $Z_{N/2}$, as both $\sigma_i$ and $S_{Higgs}\sim \sum_{\corr{ij}}U_{ij}^{N/2}$ are invariant under $\varphi_{ij}\rightarrow \varphi_{ij}+\frac{4\pi}{N}$. 

Now in order to make the remaining degrees of freedom explicit and as remarked earlier, it is consistent to assign different core energies $K$ for the $Z_N$ disclinations. Separating the $Z_{N/2}$ configurations in the gauge fields as
\begin{align}
\varphi_{ij} = \frac{2\pi}{N}m_{ij} = \frac{2\pi}{N}(2 l_{ij} + k_{ij}),
\end{align}
where $l_{ij} = 0,1,\dots, \frac{N}{2}-1$ and $k_{ij} =0,1$, we can adjust the gauge field term 
\begin{align}
K\sum_{\Box} \cos (\varphi_{\Box}) \to \nonumber \\ 
\sum_{\Box} \delta_{k_{\Box},0}K_0\cos (\frac{4\pi}{N} l_{\Box})
+ \delta_{k_{\Box},1}K_1\cos(\frac{2\pi}{N}(2 l_{\Box}+k_{\Box})) \\
= S_{Z_{N/2}}+ \sum_{\Box} \delta_{k_{\Box},1}K_1\cos(\frac{2\pi}{N}(2 l_{\Box}+k_{\Box}))
\end{align}
which is just the $Z_N$ character or conjucagy class expansion of the element $\varphi_{\Box}$. Clearly we can have independent gauge dynamics for the two $Z_{N/2}$ subgroups of $Z_N = Z_{2} \times Z_{N/2}$.

The symmetry of the Higgs ordered phase is readily apparent when we dualize the theory with the additional Higgs term Eq.\eqref{Higgs} in the unitary gauge 
\begin{align}
 \tilde{S}_N &= -\frac{1}{8\pi^2J} \sum_{\tilde{\square}} A^{2}_{\tilde{\square}} - i \sum_{\tilde{i},\mu (\Box) } A_{\tilde{i},\mu}(J^{XY}_{\Box} +\frac{2k_{\Box}}{N}+\frac{l_{\Box}}{N}) \nonumber \\ &-\sum_{\square} K_0\delta_{l_{\Box},0}\cos{(4\pi k_{\square}/N)} \\& - \sum_{\square} K_1\delta_{l_{\Box},1} \cos (2\pi (2k_{\Box}+l_{\Box})/N) \nonumber\\ &- M\sum_{\corr{ij}} \cos (\pi(2k_{ij}+l_{ij})) ~.\nonumber 
\end{align}
In the limit $M$ large, the gauge field completely freezes to the $Z_{N/2}$ sector. As a result, one obtains a $O(2)/Z_{N/2}$ gauge theory and the factor in the mutual gauge coupling term attains a value of $2/N$ instead of the original charge $1/N$ for the $Z_N$. Hence the Higgs term Eq.\eqref{Higgs} indeed effectively drives a phase transition from $C_{N}$ to $C_{N/2}$ nematic as a function of $M$ and the transition is in the Ising universality class.

To see this, note that in effect we have a "$Z_2/Z_2$" gauge theory for $\sigma_i$, although the $Z_2 $ gauge field coupling to $\sigma_i$ is of course the original $Z_N$ gauge field in the system. The phase diagram for such theories was discussed in Ref. \onlinecite{Fradkin1979}, where it was shown that the phase transition to the Higgs phase as a function of $M$ is given by the $Z_2$ Ising transition.

Such extra fields discrete fields $\sigma_{i}$ arise from some other degrees of freedom system in the original system, for example the "valley" symmetry of $A$-$B$ sublattices on the honeycomb lattice. In the disordered phase (i.e. no "valley" symmetry breaking), we can integrate out $\sigma_{i}$ in Eq. \eqref{Higgs}  to obtain an invariant term $U_{ij}^N$ that is irrelevant for the $Z_N$ gauge theory. In contrast, in the symmetry breaking phase of the $\sigma_i$ field (that arises spontaneously or explicitly) gives rise to the Higgs term Eq. \eqref{Higgs}.  Hence, the $Z_N$ to $Z_{N/2}$ phase transition is indeed analogous to the order-disorder phase transition of the Ising gauge theory.  Similarly, the other possible phase transitions arise in the same way, e.g. the transition $C_6$ to $C_2$, may be described in the same way by a Higgs terms with $N/3$ charged $Z_3$ matter. This transition is then described by a $Z_3$ Potts model \cite{Fradkin1979, Wu1982}. We can also break the $C_N$ symmetry of the nematic completely by adding a $Z_N$ Higgs field with the fundamental charge.

There is also the possibility of topological phase transitions, e.g. between the $Z_{N}$ deconfined and the $Z_{N/2}$ deconfined phase, by tuning the gauge coupling $K_1 \to 0$. More generally, a transition can be tuned in terms of the gauge couplings $\{K_i\}_{i\in Z_N}$ for a subgroup of $Z_N$ in the $Z_N$ deconfined phase of the gauge theory, leading to Ising or Potts transitions to the deconfined phase of the subgroup. Similarly, for example in the limit of small $K_1$ above, there is a phase transition between the $C_N$ nematic and the deconfined $Z_{N/2}$ phase as function of $J$. Admittely, the tuning of $K_1$ independently is in both cases physically somewhat artificial. Note that the condensation of the \emph{odd} $Z_{N}$ fluxes and the $2\pi$-vortices, as required for the $Z_{N/2}$ deconfined phase, will always disorder the matter field. Thus the latter transition will actually involve an $XY$ transition of the matter fields plus a confinement transition for the \emph{odd} $Z_N$ fluxes/vortices. In this particular case, one would expect an Ising or Potts ($Z_2$ or $Z_3$) phase transition for the gauge fields. The left over $Z_{N/2}$ gauge degrees of freedom are then in the deconfined phase once the matter field disorders.

\section{conclusion and discussion}
In this paper, we have provided a full symmetry classification of quantum nematic order in 2+1 dimensions by dislocation melting of crystalline phases We further constructed an $O(2)/Z_N$ gauge theory describing the nematic phases in terms of two parameters: the nematic interaction $J$ and a defect suppression term $K$, related to the gauge fields. The resulting phase diagram contains at least three different phases: the $C_N$ nematic phase, isotropic liquid and a topological phase arising from the gauge fields. 

Using our gauge theory description, we can further generically describe all the universal properties of the possible $C_N$ nematic phases, in particular the various phase transition between the $C_N$ nematics in addition to the nematic-to-isotropic liquid phase transitions. This is due to the efficient way the introduced auxiliary gauge degrees of freedom encode for the desired nematic symmetries.

We also verified the salient points of the phase diagram of our gauge model with Monte Carlo simulations.

In addition to the conventional nematic phases, we have shown how the theory can be applied beyond the continuous symmetry breaking scheme of nematic ordering. This amounts to taking into account the gauge degrees of freedom as independent degrees of freedom. We found "deconfined" topological phases corresponding each $C_N$ nematic phase, similar to that of Ref. \onlinecite{Lammert1995}. In these phases the gauge degrees of freedom themselves play a central role and there is no long range nematic order. Conceptually these are two-dimensional analogues of the spin nematic phase and are similar to those arising in the "deconfined" quantum criticality scenario \cite{Senthil2004deconfined}. In particular, the topological phase is separated from the nematic phase by a second order transition (of the nematic degrees of freedom), although the matter and gauge fields both go through a phase transition and the behavior of the gauge fields is only revealed by a non-local string order parameter.

The strategy for the classification of nematic phases via melting and point groups in 2d can be generalized to the 3d case.  Descending from the 230 space groups of 3d crystals, it follows within the same consideration that the nematic phases are characterized by the 32 crystalline subgroups of $O(3)$, i.e. the three dimensional point groups. The non-abelian nature of these groups makes the generalization to three dimensions fundamentally different. For instance, the analog of our $O(2)/Z_N$ lattice would be an $O(3)$ matter field coupled to a non-abelian discrete gauge field and thus considerably more involved. These issues will therefore be addressed in future work.

\textit{Acknowledgements.} We thank Vladimir Cvetkovic and Aron J. Beekman for useful discussions and Referee II for the valuable remarks. This work was supported by the Netherlands foundation for Fundamental Research of Matter (FOM). Ke Liu is supported by the State Scholarship Fund program organized by China Scholarship Council (CSC).

\appendix

\section{Imaginary time formalism}\label{App:imaginarytime}
As is well-known, the quantum statistical problem with Hamiltonian $H$ at inverse temperature $\beta=1/T$ ($k_B\equiv 1$) reduces to classical field theory in three Euclidean dimensions, with the imaginary time action $S$ and periodic imaginary time $\tau \simeq \tau + \beta$ ($\hbar\equiv 1$). In this paper, we will solely focus on the $T=0$ quantum phase transitions of the nematic phases described by our gauge model Eqs. \eqref{Eq:XY-action}, \eqref{eq:Zn-action}. Since this model is based on the introduction of the gauge field degrees of freedom relating to the spatial symmetries of the nematic, we now clarify their role in our imaginary time action.

The $N=1$ case is the familiar quantum $XY$ model in 2+1 dimensions. This has the Hamiltonian
\begin{equation}
H_{XY}[J_0, J_1] = \frac{1}{2J_0} \sum_{\vec{x}} L_{\vec{x}}^2 + J_1 \sum_{\vec{x}, \vec{i}} \cos (\theta_{x} - \theta_{\vec{x}+\vec{i}}), \label{eq:quantumXY}
\end{equation}
where $L_{\vec{x}}$ is the two-dimensional angular momentum canonically conjugate to the rotor field, $[L_{\vec{x}}, \theta_{\vec{y}}] = \im \delta_{\vec{x},\vec{y}}$. Here the $\vec{x}$ label spatial lattice sites and $\vec{i}$ spatial unit vectors. The imaginary time formulation relates this to the Euclidean action
\begin{align}
S_{XY}[J_0, J_1] =& \nonumber\\
\int^{\beta}_0\de\tau \sum_{\vec{x}}& \frac{J_0}{2} (\doo_{\tau} \theta_{\vec{x}}(\tau))^2  + J_1 \cos (\theta_{x}(\tau) - \theta_{\vec{x}+\vec{i}}(\tau)), \nonumber
\end{align}
which is the highly anisotropic limit ($J_0 \to \infty$ and $a_{\tau} \to 0$ with $J_0a_\tau=$ const.) of
\begin{align} \label{eq:classicalXY}
S_{XY} =&  \int^{\beta}_0 \de \tau J_0 \sum_{\vec{x}, \tau} \cos (\triangle_{\tau} \theta_{\vec{x}}) + J_1 \sum_{\vec{x},\vec{i}} \cos (\theta_{x} - \theta_{\vec{x}+\vec{i}})  \nonumber \\
\sim& J \sum_{\corr{ij}} \cos (\theta_i - \theta_j).
\end{align}
By the standard lore of field theory and critical phenomena, the isotropic model described by Eq. \eqref{eq:classicalXY} and its particular limit Eq. \eqref{eq:quantumXY} describing the 2+1-dimensional quantum system in the operator formalism are expected to carry the same universal properties. This justifies the analysis of the latter model Eq. \eqref{eq:classicalXY} with regards to the quantum system. Nevertheless, the quantum model is really described by the rotational symmetries in the spatial dimensions, and the full three dimensional isotropy of Eq. \eqref{eq:classicalXY} broken by the (periodic) imaginary time direction. Restoring units, we see that $J_0 \sim J_0/\hbar^2$, and this sets the size of the quantum fluctuations in the system.

Note in particular that the quantum model features the two-dimensional $XY$ model at every constant $\tau$-slice, but it is the proliferation of the time-like vortex loops of arbitrary length in the imaginary time direction that drives the phase transition, and leads to the similar critical behavior as in the classical model. On the other hand, in the extreme high-temperature limit $\hbar\beta\to 0$ the quantum model reduces to the classical two-dimensional $XY$ model.

For the models $O(2)/Z_N$, it is more instructive to start with the imaginary time actions in Eqs. \eqref{Eq:XY-action}, \eqref{eq:Zn-action}, the highly anisotropic limits of which are
\begin{align}
S_I \sim \int^{\beta}_0\de \tau~ & \frac{J_0}{2} \sum_{\vec{x}} (\triangle_{\tau} \theta_{\vec{x}} +\phi_{\vec{x}})^2  \nonumber\\
&+ J_1 \sum_{\vec{x},\vec{i}} \cos (\triangle_{\vec{i}}\theta_{\vec{x}}+\varphi_{\vec{x},\vec{i}}) \nonumber \\
S_G \sim  \int^{\beta}_0\de \tau~ & \sum_{\vec{x},\vec{i}} \frac{K_0}{2} (\triangle_{\tau}\varphi_{\vec{x},\vec{i}}+\phi_{\vec{x}+\vec{i}} -\phi_{\vec{x}})^2 \nonumber \\
&+ K_1 \sum_{\Box_{\vec{x}}} \cos (\varphi_{\Box_{\vec{x}}}) \nonumber
\end{align}
where $\Box_{\vec{x}}$ label the spatial plaquettes and we have denoted $\phi_{\vec{x}} \equiv \varphi_{\vec{x},\tau}$ the time component of the gauge potential. The gauge transformations are given by
\begin{align}
\phi_{\vec{x}} &\to \phi_{\vec{x}}+\frac{2\pi}{N}\triangle_{\tau}\lambda_{\vec{x}}\\
\varphi_{\vec{x},\vec{i}}&\to \varphi_{\vec{x},\vec{i}} + \frac{2\pi}{N}\triangle_{\vec{i}}\lambda_{\vec{x}}
\end{align}
where $\lambda_{\vec{x}}(\tau)$ is an arbitrary integers mod $Z_N$ valued function on the lattice $\{\vec{x},\tau\}$.

These lead to the Hamiltonians
\begin{align}
H_I = \frac{1}{2J_0} \sum_{\vec{x}} \Pi^2_{\vec{x}} + J_1 \sum_{\vec{x},\vec{i}} \cos (\triangle_{\vec{i}}\theta_{\vec{x}} + \varphi_{\vec{x},\vec{i}}) \\
H_G = \frac{1}{2K_0} \sum_{\vec{x},\vec{i}} E^2_{\vec{x},\vec{i}} + K_1 \sum_{\Box_{\vec{x}}} \cos(\varphi_{\vec{x},\vec{i}}), \label{eq:Hnematic}
\end{align}
where $\Pi_{\vec{x}} = J_0 (\triangle_{\tau}\theta_{\vec{x}} +\phi_{\vec{x}})$ is the canonical momentum of the gauge coupled rotor, including the time component of the gauge potential $\phi_{\vec{x}}$. Similarly, $E_{\vec{x},\vec{i}} = K_0 (\triangle_{\tau}\varphi_{\vec{x},\vec{i}} + \triangle_{\vec{i}}\phi_{\vec{x}})$ is the $Z_N$ electric field, canonically conjugate to the gauge potential, i.e. $[E_{\vec{x},\vec{i}}, \varphi_{\vec{y},\vec{j}}] = 2\pi \im /N \delta_{x+\vec{i},\vec{y}+\vec{j}}$. For a gauge system, the canonical formalism necessarily specifies a gauge and associated constraints. The above form of the Hamiltonian, where the field $\phi_{\vec{x}}$ appears without time derivates, is valid in the gauge where we set
\begin{equation}
\phi_{\vec{x}}(\tau) = 0 \textrm{ for all } \vec{x}.
\end{equation}
This eliminates the gauge transformations $\lambda_{\vec{x}}(\tau)$ that depend on the $\tau$ direction on the lattice. However, we still have satisfy the constraint
\begin{equation}
\frac{\delta H}{\delta \phi_{\vec{x}} } = 0 = \sum_{i} \triangle_{\vec{i}}E_{\vec{x},\vec{i}} - Q_{\vec{x}},
\end{equation}
which is Gauss' law. The charge is defined in terms of the rotor-field as
\begin{equation}
Q_{\vec{x}}= \Pi_{\vec{x}} = J_0\triangle_{\tau}\theta_{\vec{x}},
\end{equation}
i.e. the rotor angular momentum. The remaining gauge degrees of freedom are determined by transformations of the form
\begin{align}
\varphi_{\vec{x},\vec{i}} &\to \varphi_{\vec{x},\mu}+2\pi \triangle_{\vec{i}}\lambda_{\vec{x}}/N.
\end{align}
where $\lambda_{\vec{x}}$ is an arbitrary integer mod $Z_N$ on the spatial lattice $\vec{x}$ but constant in $\tau$. We conclude that the Hamiltonian Eq. \eqref{eq:Hnematic} has only spatial gauge symmetry, as is appropriate for the quantum nematics.

\section{Villain approximation and duality of the $XY-Z_N$ model}\label{XYN Duality}

We now briefly recollect the Villain approximation and the $XY$-duality transformation for the theory Eq.\eqref{matter action}\cite{Savit1977, Senthil2000}. For more details we refer the reader to e.g. Ref. \onlinecite{Savit1977}. In the following we will use vector notation on the lattice as $ L_{i,\mu}\equiv L_{ij}$, which represents the link variable $L_{ij}$ on the link $i,\mu \equiv ij$ from $i$ to $j=i+\mu$ in the direction of the unit vector $\vec{e}_{\mu}$ ($\mu=x,y,\tau$ in 2+1 dimensions). We will also denote with $\triangle_{\mu}$ the finite difference operator $\triangle_{\mu}f(i) \equiv f(i+\mu)-f(i)$.

The Villain approximation, valid in the limits $J\to\infty$ and $J\to 0$, takes the form
\begin{align}
e^{J \cos\Theta_{i,\mu}} &\to 
N_V(J) \sum_{l_{i,\mu}\in \integers} e^{-\frac{J_V}{2}(\Theta_{i,\mu} + 2\pi l_{i,\mu})^2}\\
&= N_V(J) \sum_{L_{i,\mu}\in\integers} e^{-L^2_{i,\mu}/(2J_V)} e^{iL_{i,\mu}\Theta_{i,\mu}},\label{Villain}
\end{align}
where $l_{i,\mu}$ and $L_{i,\mu}$ are integer valued auxiliary fields, $J_V(J)$ is the effective Villain temperature and $N_V(J)=\sqrt{2\pi J}I_0(J)$ is an analytic normalization factor \cite{Janke1986, Jose1977, Villain1975}. Henceforth we will simply denote the effective coupling $J_V$ as $J$. 
\subsection{Duality}
We apply Eq. \eqref{Villain} to $\Theta_{i,\mu}=\triangle_{\mu}\theta_i+\varphi_{i,\mu}$ to dualize the $XY$ variables $\theta_i$ \cite{Savit1977}:
\begin{align}
e^{-S_I}&=e^{J\sum_{i,\mu}\cos(\triangle_{\mu}\theta_i+\varphi_{i,\mu})}
\nonumber \\
&\rightarrow \sum^{\infty}_{L_{i,\mu}=-\infty}
e^{\sum_{i,\mu}[-L^2_{i,\mu}/(2J) + iL_{i,\mu}(\triangle_{\mu}\theta_i+\varphi_{i,\mu})]}.
\end{align}
We rewrite the sum over the $L_{i,\mu}$ as
\begin{align}
\sum_{L_{i,\mu} = -\infty}^{\infty} = \int \mathcal{D}[L_{i,\mu}] \sideset{}{'}\sum_{V_{\mu,i} = -\infty}^{\infty} e^{\sum_{i,\mu} 2\pi\im L_{i,\mu}V_{i,\mu}}, 
\end{align}
which loosely speaking takes into account the vortices by the substitution $\triangle_{\mu}\theta_i \to \triangle_{\mu}\theta_i + 2\pi V_{i,\mu}$, where now $-\infty < \theta_i < \infty$ and the integers $V_{i,\mu}$ are related to the local vortex density as $J^{XY}_{\Box} = \sum_{\nu\lambda\in\Box}\epsilon_{\mu\nu\lambda}\triangle_{\nu}V_{i,\lambda} = V_{\Box}$. Moreover, the action for a configuration depends only different vortex numbers $J^{XY}_{\Box}$, and the tilde in the sums over $V_{i,\mu}$ in the partition function refers to a constraint to eliminate the overcounting\cite{Savit1977}.

Now we can integrate over the rotors $\theta_{i}$ in the partition function
$Z=\sum_{\{\varphi_{i,\mu}\}} \int \mathcal{D}[\theta_i] e^{-S}$ to get
\begin{align*}
\int \mathcal{D}[\theta_i]e^{i\sum_{i,\mu}\triangle_{\mu}L_{i,\mu}\theta_i}
=& \prod_{i} \int^{\infty}_{-\infty} d\theta_i e^{i\sum_{\mu}\triangle_{\mu}L_{i,\mu} \theta_i}\\
=& \prod_{i} 2\pi \delta(\sum_{\mu}\triangle_{\mu}L_{i,\mu}).
\end{align*}

The constraint $\sum_{\mu}\triangle_{\mu}L_{i,\mu}=0$ is the discrete version of $\nabla\cdot \vec{L}=0$ of a vector field $\vec{L}=L_{\mu}$. This can be solved by introducing a field $A_{\tilde{i},\mu}$ on the dual lattice satisfying
\begin{equation}\label{dual fields}
2\pi L_{i,\mu} =\epsilon_{\mu\nu\lambda} \triangle_{\nu}A_{\tilde{i},\lambda} =A_{\tilde{\square}},
\end{equation}
where $\tilde{i}$ denotes the dual lattice site and $A_{\tilde{\square}}$ is the dual plaquette pierced by the bond $i,\mu$ on the original lattice. However, shifting $A_{\tilde{i},\mu}$ by
\begin{equation}
A_{\tilde{i},\mu} \to A_{\tilde{i},\mu} + \triangle_{\mu}a_{\tilde{i}},
\end{equation}
where $a_{\tilde{i}}$ is an arbitrary real function on the dual lattice, leads to the same $L_{i,\mu}$ and therefore to physically equivalent configurations. In terms of $A_{\tilde{i},\mu}$ this represents a gauge symmetry. In fact, this ambiguity leading to overcounting is exactly similar to that arising in terms of $V_{i,\mu}$.

The theory now can be re-expressed using  Eq. \eqref{dual fields},
\begin{align}\label{dual S}
\tilde{S}_N = -\frac{1}{8\pi^2J} \sum_{\tilde{\square}} A^{2}_{\tilde{\square}} - i \sum_{\tilde{i},\mu (\Box) } A_{\tilde{i},\mu}(J^{XY}_{\Box} +\frac{\varphi_{\Box}}{2\pi}) \nonumber \\ -K\sum_{\square}\cos{(\varphi_{\square})}.
\end{align}
and where we sum over the vortex numbers $J^{XY}_{\Box}\in\integers$, the non-compact dual gauge field $A_{\tilde{i},\mu}\in 2\pi\real$, and the original $Z_N$ gauge field $\varphi_{ij}$ in the partition function. 

The first term represent a non-compact $U(1)$ gauge theory for the dual field $A_{\tilde{i},\mu}$, with the summation $\sum_{\tilde{\square}}$ running over  the dual lattice plaquettes. The third term is just the original $Z_N$ gauge theory. The vortex densities $V_{\Box}$ and $\varphi_{\Box}$ are both charged under the dual gauge field $A_{\tilde{i},\mu}$, as in the usual $XY$-duality. Moreover, noting that $\varphi_{ij} = \frac{2\pi}{N}m_{ij}$, the charge coupling is
\begin{equation}
\im \sum_{\tilde{i},\mu(\Box) } A_{\tilde{i},\mu}(J^{XY}_{\Box} + \frac{1}{N}m_{\Box}),
\end{equation}
whence the $Z_N$ vortices carry a fractional charge of $1/N$ as compared to the $2\pi$-vortices, exactly as we would expect. The dual gauge symmetry dictates that
\begin{equation}
\sum_{\mu} \triangle_{\mu} (J^{XY}_{\Box}+\frac{\varphi_{\Box}}{2\pi}) = 0.
\end{equation}
However, the $Z_N$ field strength $\varphi_{\Box}$ is conserved only up to integers
\begin{align}
\triangle_{\mu}\varphi_{\Box} = 0 \mod 2\pi
\end{align}
and therefore can source the integer current $J^{XY}_{\Box}$. This means that $2\pi$-vortex lines can begin/end on sites where $Z_N$ fluxes end/begin and is a consequence of the compactness of the $Z_N$ gauge group \cite{Polyakov1975} in combination with the usual $XY$ vortices. Intuitively this is clear in the sense that we can consider the vortices $J_{\Box}^{XY}$ as $N$-tuples of the $Z_N$ fluxes. We also see that the total defect current $J^{XY}_{\Box}+\varphi_{\Box}/2\pi$, or the defect charge in units of $1/N$, is conserved. Since the defects interact with Coulomb forces in the ordered phase, just as in the usual $XY$ model, such "splitting" is energetically costly and the main contribution comes from closed defect loops of $J^{XY}_{\Box}$ and $\varphi_{\Box}$.

We conclude that the dual theory is given by a non-compact $U(1)$ gauge theory coupled to the original $Z_N$ gauge theory, with the coupling term encoding the mutual statistics of $2\pi/N$ between the $Z_N$ flux and the original rotors $\theta_{i}$, whose density is represented by the flux $A_{\tilde{\Box}}$. In addition to the $Z_N$ vortices, the standard $2\pi$-vortices $J_{\Box}^{XY}$ are charged under the $U(1)$ dual gauge field, as in the $XY$-duality. Note that while the $Z_N$ vortices carry the core energy $K$, the usual core energy for the $2\pi$-vortices $J^{XY}_{\Box}$ has not been explicitly included. In this degenerate limit, the summation over $J_{\Box}^{XY}$ can be performed leading to the constraint $A_{\tilde{i},\mu} \in 2\pi\integers$. This in effect creates a mass gap in the system and is the crude analogue of the usual Higgs symmetry breaking and mass in the $U(1)$ gauge theory that occurs in the $J$ small regime.

\section{The deconfined phase}

Here we present a more detailed discussion about the deconfined phase at large $K$. First we define the string order parameter for the topological phase and then focus on the $K\to\infty$ limit of the theory.  

\subsection{Fredenhagen-Marcu order parameter}

In terms of usual Landau symmetry breaking arguments, one might be inclined to think that the field $e^{iN\theta}$ constitutes an order parameter for any phase transition driven by the coupling $J$ involving the $C_N$ ordered nematic. However in the large $K$ regime, this order parameter is not actually sufficient to fully characterize the phase transition due to the presence of the $Z_N$ gauge fields. For the pure $Z_N$ gauge theory an order parameter is given by the Wilson loop
\begin{align}\label{Wilson loop}
W(C_L)=\langle\prod_{\langle ij\rangle \in C}U_{ij}\rangle,
\end{align}
where $C_L$ denotes a closed path of length $L$ on the lattice. As is well-known, the $L\to\infty$ asymptotics of characterize the confinement-deconfinement transition in a pure gauge theory. However, it in general fails to do so in the presence of any charged matter fields. Hence, it would be worthwile to identify an order parameter that could distinguish the order-disorder for the matter field {\it and } the confinement-deconfinement for the gauge field simultaneously.

Inspired by the string operator in matter-coupled lattice gauge theory \cite{Kogut1975} and the string correlator recently suggested for systems with topological matter\cite{Cobanera2013} , we can define a string order parameter including both rotor fields and gauge fields as
\begin{align}
\mathcal{O}(C_{L/2})=\langle n_k^{*}\left(\prod_{ij\in C_{L/2}(k,m)}U_{ij}\right)n_m\rangle \label{string1}
\end{align}
where $C_{L/2}(k,m)$ is an arbitrary path of length $L/2$ connecting two rotors $n_k$ and $n_m$ . It is straightforward to see that this string order parameter is invariant under the gauge transformation $\theta_k\rightarrow \theta_k+\frac{2\pi}{N}, \varphi_{ij}\rightarrow\varphi_{ij}-\frac{2\pi}{N}$. We can renormalize the string order parameter \eqref{string1} with the Wilson loop \eqref{Wilson loop} to obtain the Fredenhagen-Marcu order parameter\cite{Fredenhagen1986,Fredenhagen1988}
\begin{align}
R(C_L)\equiv\frac{\mathcal{O}(C_{L/2})}{\sqrt{W(C_L)}}=\frac{\langle n^*_k\left(\prod_{ij\in C_{1/2}}U_{ij}\right)n_m\rangle}{\sqrt{\langle\prod_{ij\in C}U_{ij}\rangle}}.
\end{align}
It can be shown \cite{Fredenhagen1986, Fredenhagen1988, Gregor2011} that $R(C_L)$ distinguishes the $Z_N$ deconfined phase from the $C_N$ nematic phase and the isotropic liquid:
\begin{align}
\lim_{L\rightarrow\infty}R(C_L)&=0 \quad \text{$Z_N$ deconfined phase}\nonumber\\
\lim_{L\rightarrow\infty}R(C_L)&\neq0 \quad \text{$C_N$ nematic or isotropic liquid}. \nonumber
\end{align}

The difference between the small $K$ and large $K$ limits of $R(C_L)$ is due to the behavior of the gauge field in the presence of the matter field. In the small $K$ limit, the gauge field is strongly fluctuating which renders isolated $Z_N$ defects absent, in analogy to the confined phase in pure gauge theory. However, due to the matter fields, a phase with free $Z_N$ defects is possible for large enough $J$ but this transition is driven by the matter field. Actually, in a gauge theory with matter, any string order parameter of the gauge field always decays exponentially, which is why the denominator is introduced in $R(C_{L})$. However, only in the phase with deconfined gauge fields and disordered matter fields, the limit $L\to\infty$ results in a non-zero value and therefore serves as the correct order parameter of the topological phase.

\begin{figure}[!tp]
\centering
\includegraphics[width=0.3\textwidth]{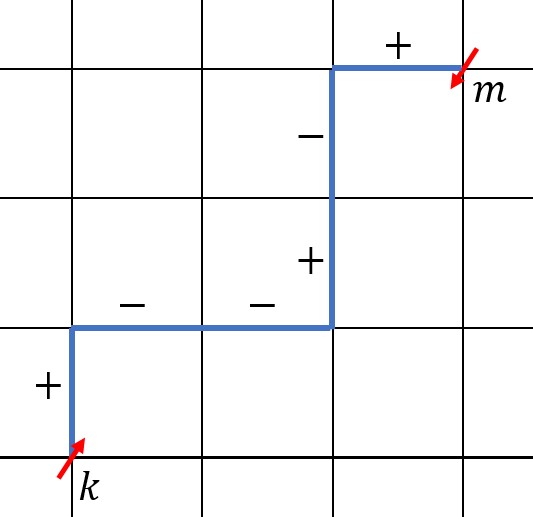}
\caption{ A gauge invariant string order parameter $\mathcal{O}(L)$ defined by Eq.\eqref{string1}. Red arrows are rotor fields at the ends of the string. Blue bonds indicate an arbitrary gauge string that connects the two rotors. Take the $O(2)/Z_2$ case as an example, the gauge field living on each bond can take two values, i.e. $+1$ and $-1$, under the constraint $U_{\square}=1$. }
\label{Fig: string}
\end{figure}

\subsection{The $K\rightarrow\infty$ limit: phase transition and order parameter}\label{sec:Klarge}

In the $K\rightarrow\infty$ limit, only defect free configurations are allowed. Hence all non-trivial plaquette excitations of the $Z_N$ gauge fields are prohibited, leading to the constraints $\varphi_{\square}=0$ or $U_\square=1$ on all plaquettes for the $Z_N$ lattice gauge fields $U_{ij}$. This constraint allows us to parameterize the gauge field as $U_{ij}=u_i^{*}u_j$, where $u_i=e^{i z_i}$ and $z_i$  are $Z_N$ fields defined on the lattice sites $i$ as $z_i=\frac{2\pi m_i}{N}$ with $m_i=0,1,...,N-1$.  As a consequence, the denominator in $R(L)$ equates to unity and $R(L)$ reduces to the string order parameter $\mathcal{O}(L)$.

The underlying reason for the above results is that the partition function of the $O(2)/Z_N$ turns out to be equivalent to that of the $XY$ model in the $K\rightarrow\infty$ limit. This can be shown directly. To this end, we apply the constraint $U_{ij}=u_i^* u_j$ to rewrite Eq.  \eqref{Eq:action} in the following form
\begin{equation}\label{SKinf}
S_{K\rightarrow\infty}=-J\sum_{\langle ij\rangle}[(u_in_i)^{*}(u_jn_j)+c.c].
\end{equation}
As a result, the partition function becomes
\begin{align}\label{SKinf1}
Z_{K\rightarrow\infty}&=\sum_{\{u_i\}}\int_{0}^{2\pi} \mathcal{D}[\theta_i] e^{-S_{K\rightarrow\infty}}
\nonumber\\
&=\sum_{\{u_i\}}\int_{0}^{2\pi} \mathcal{D}[\theta_i] e^{J\sum_{ij}[(u_in_i)^{*}(u_jn_j)+c.c]}.
\end{align}
Now we can shift the variables $n_i \to n_i' \equiv u^*_i n_i$ at every site $i$. 

By gauge invariance of the action and the measure $\mathcal{D}[\theta_i] = \prod_{i} d\theta_i$, we get
\begin{align}
\label{FF}
\sum_{u_i}\int_{0}^{2\pi} d\theta_i\mathcal{F}(u_in_i) = N \int_{0}^{2\pi} d\theta_i \mathcal{F}(n_i)
\end{align}
for any arbitrary functional $\mathcal{F}$. Henceforth,
\begin{align}
\label{duality_K_large}
Z_{K\rightarrow\infty}=N^{N_s}\int_{0}^{2\pi} \mathcal{D}[\theta_i] e^{J\sum_{ij}[n_i^* n_j+c.c]},
\end{align}
where $N_s$ is the number of the lattice sites. The above form makes it immediately apparent that the partition function is just the partition function of the $XY$ model up to a constant prefactor, i.e. $Z_{K\rightarrow\infty}=N^{N_{site}}Z_{XY}$.

However, the usual $XY$ field $n_i=e^{\im \theta_i}$ is not gauge invariant and therefore cannot characterize the phase transition as an order parameter. Instead, we need a gauge invariant quantity e.g. $(n_i)^N=e^{iN\theta_i}$, which is a composite field in the usual $XY$ model. The correlation function of $(n_i)^N$ can be written as:
\begin{align}
\langle (n_k^*)^N (n_m)^N\rangle_{S_N}=\langle e^{iN(\theta_k-\theta_m)}\rangle_{XY} \label{eq:Ntheta correlator}
\end{align}
For the physical matter field $n^N_i$, the phase transition is in the so-called $XY^{\star}$ universality class. Put differently, this is just the statement that while the partition function \eqref{duality_K_large} is exactly same as that of the $XY$ model, the relevant correlation function at the transition is a composite field rather than the usual $XY$ field. This affects some aspects of the universality class of the transition, for instance the anomalous dimension $\eta$ \cite{Grover2010, Isakov2012}. 

Similar arguments apply to any gauge invariant field in the model. Let us now discuss this in more detail. First define a gauge invariant ``bond'' $Z_{ij} \equiv n_{i}^{*} U_{ij} n_{j}$ \cite{Zohar2005}.

The most general gauge invariant average is then given by $\langle \mathcal{F}(\{Z_{ij}\})\rangle$ where $\mathcal{F}$ is an arbitrary functional of the bond variables $Z_{ij}$ on the lattice. By virtue of Elitzur's theorem, the expectation values of all other quantities must vanish.
A particular such average is
\begin{eqnarray}
\label{ops}
\langle \prod_{\langle ij \rangle \in C} Z_{ij}^{p_{ij}} \rangle
\end{eqnarray}
where $\mathcal{F}$ is defined by the set of integers $\set{p_{ij}}_{\corr{ij}\in C}$ along some path $C$. In order to compute the expectation value, the action may be reformulated as
\begin{eqnarray}
\label{SN}
S_{N} = &&- \frac{1}{2} \sum_{\langle i j \rangle} J_{ij} (Z_{ij} + Z_{ij}^{*}) \nonumber
\\ &&- \frac{1}{2} \sum_{\{ ijkl \} \in \Box} K_{\Box} (Z_{ij} Z_{jk} Z_{kl} Z_{li} + c..c).
\end{eqnarray}
In Eq. (\ref{SN}), we generalized the theory defined in Eqs. (\ref{matter action}, \ref{gauge action}) to allow for locally varying coupling terms $J_{ij}$ and $K_{\Box}$.
Note that Eq. (\ref{ops}) reduces to string correlator $\mathcal{O}(C_{L/2})$ in Eq. (\ref{string1}) when the integers $p_{ij}$ are taken unity along the path $C$.
Now, here is a simple yet important point:
\begin{eqnarray}
\mathcal{O}(C_{L/2}) = \Big[ \prod_{ij \in C} \frac{\delta}{\delta J_{ij}} \Big] \ln Z|_{J_{ij} = J, K_{\Box}=K},
\end{eqnarray}
with $Z$ the partition function.
However, as we have proven, $Z$ becomes the partition function of the XY model in the $K\rightarrow\infty$ limit and this result generalizes for non-uniform couplings $J_{ij}$. It follows that
\begin{eqnarray}\label{eq: SO}
\Big[ \prod_{ij \in C} \frac{\delta}{\delta J_{ij}} \Big] \ln Z(K\to\infty)|_{J_{ij} = J} = \langle n_{k}^{*} n_{m} \rangle_{XY}.
\end{eqnarray}
We thus observe that the string order parameter reduces to the standard two-point correlator of the XY model in $K\rightarrow\infty$ limit.
This, of course, is also seen by evaluating $\mathcal{O}(C_{L/2})$ in the specific gauge $U_{ij} =1$ for all $ij$.

Repeating, {\it mutatis mutandis}, the above steps, it is also readily seen that, for any integer $p$, the average
\begin{eqnarray}\label{eq: SOp}
\langle (n_k^{*})^{p}\left(\prod_{ij\in C_{k,m}}U_{ij}^{p}\right)(n_m)^{p} \rangle_{S_{N}}
= \langle (n_{k}^{*} n_{m})^{p} \rangle_{XY}.
\end{eqnarray}
This is the generalization of Eq.\eqref{eq:Ntheta correlator}, since the gauge field string becomes trivial when $p=N$ and drops out from the right hand side.

At last, the astute reader may note that the above steps in Eqs. (\ref{SKinf})-(\ref{duality_K_large}) can easily be generalized to allow for varying couplings constants $J_{ij}$. This then leads to the equivalence of the matter coupled gauge theory of Eq. (\ref{Eq:action}) in the limit $K\to \infty$
to  that of the $XY$ model not only for the the standard uniform $XY$ model, but also for the far richer case of arbitrary couplings.
Yet another illuminating way to obtain this result for general couplings $J_{ij}$
is obtained by examining the gauge invariant formulation of the action in Eq. (\ref{SN}).
We note that in the $K \to \infty$ limit, the product
\begin{eqnarray}
\label{product1}
\prod_{ij \in \Box} Z_{ij} =1,
\end{eqnarray}
for any plaquette $\Box$ on the lattice.
The action of the XY model generalized for arbitrary couplings $J_{ij}$ then becomes
\begin{eqnarray}
\label{xyz}
S_{XY}=  && - \frac{1}{2} \sum_{ij} J_{ij} (n_{i}^{*} n_{j} + c.c.) \nonumber
\\ = && - \frac{1}{2} \sum_{ij} J_{ij} (Z_{ij} + Z_{ij}^{*})|_{\prod_{ij \in \Box} Z_{ij} =1}.
\end{eqnarray}
Where in the second line of Eq. (\ref{xyz}), the bonds $Z_{ij}$ are subject to the condition of Eq. (\ref{product1}).
This is so as the product around any closed loop of the interactions in the XY model must satisfy Eq. (\ref{product1}). That is, around any plaquette
\begin{eqnarray}
(n_{i}^{*} n_{j})(n_{j}^{*} n_{k}) (n_{k}^{*} n_{l}) (n_{l}^{*} n_{i}) =1.
\end{eqnarray}
On the other hand, the expression for the $XY$ action in the second line of Eq. (\ref{xyz}) is nothing but the action of the matter coupled gauge theory in the limit of $K \to \infty$
(where the gauge action of Eq. (\ref{gauge action}) simply gives rise to the constraint of Eq. (\ref{product1})).
Putting all of the pieces together, this establishes equivalence ( an exact {\it bond algebraic duality} \cite{Zohar2011} 
of the $XY$ model with general couplings $J_{ij}$ and the $O(2)/Z_N$ theory with the same couplings in the $K \to \infty$ limit.

\bibliographystyle{apsrev}
\bibliography{2dnematic}

\end{document}